\documentclass[twocolumn]{aastex631}

\usepackage{soul}

\shorttitle{An excess of SMGs in protocluster cores}
\shortauthors{Araya-Araya et al.}
\usepackage{amsmath} 
\usepackage[normalem]{ulem} 
\usepackage{xspace}
\definecolor{darkgreen}{rgb}{0.0,0.7,0.0}

\newcommand{\msun}{~\mathrm{M_{\odot}}}

\begin{document}

\title{Modelling the multi-wavelength detection of protoclusters. I: An excess of submillimetre galaxies in protocluster cores}

\author[0000-0003-2860-5717]{Pablo Araya-Araya}
\affiliation{Departamento de Astronomia, Instituto de Astronomia, Geofísica e Ciências Atmosféricas,
Universidade de São Paulo, \\
Rua do Matão 1226, Cidade Universitária, 05508-900, São Paulo, SP, Brazil}
\affiliation{Center for Computational Astrophysics, Flatiron Institute, 162 Fifth Avenue, New York, NY 10010, USA}
\email{paraya-araya@usp.br}

\author[0000-0001-8855-6107]{Rachel K. Cochrane}
\affiliation{SUPA, Institute for Astronomy, Royal Observatory Edinburgh, EH9 3HJ, UK}
\affiliation{Columbia Astrophysics Laboratory, Columbia University, 550 West 120th Street, New York, NY 10027, USA}
\affiliation{Center for Computational Astrophysics, Flatiron Institute, 162 Fifth Avenue, New York, NY 10010, USA}

\author[0000-0003-4073-3236]{Christopher C. Hayward}
\affiliation{Center for Computational Astrophysics, Flatiron Institute, 162 Fifth Avenue, New York, NY 10010, USA}
\affiliation{Kavli Institute for the Physics and Mathematics of the Universe (WPI),  \\ The University of Tokyo Institutes for Advanced Study, The University of Tokyo, Kashiwa, Chiba 277-8583, Japan}

\author[0000-0001-9320-4958]{Robert M. Yates}
\affiliation{Centre for Astrophysics Research, University of Hertfordshire, Hatfield, AL10 9AB, UK}

\author[0000-0002-3876-268X]{Laerte Sodré Jr.}
\affiliation{Departamento de Astronomia, Instituto de Astronomia, Geofísica e Ciências Atmosféricas,
Universidade de São Paulo, \\
Rua do Matão 1226, Cidade Universitária, 05508-900, São Paulo, SP, Brazil}

\author[0000-0002-9191-5972]{Marcelo C. Vicentin}
\affiliation{Departamento de Astronomia, Instituto de Astronomia, Geofísica e Ciências Atmosféricas,
Universidade de São Paulo, \\
Rua do Matão 1226, Cidade Universitária, 05508-900, São Paulo, SP, Brazil}

\author[0000-0002-1619-8555]{Douglas Rennehan}
\affiliation{Center for Computational Astrophysics, Flatiron Institute, 162 Fifth Avenue, New York, NY 10010, USA}

\author[0000-0002-8214-7617]{Roderik Overzier}
\affiliation{Leiden Observatory, Leiden University, PO Box 9513, NL-2300 RA Leiden, The Netherlands}
\affiliation{Observat\'orio Nacional/MCTI, Rua General Jos\'e Cristino, 77, S\~ao Crist\'ov\~ao, 20921-400, Rio de Janeiro, RJ, Brazil}

\author[0000-0002-8801-4911]{Marcel van Daalen}
\affiliation{Leiden Observatory, Leiden University, PO Box 9513, NL-2300 RA Leiden, The Netherlands}

\begin{abstract}

Studies of galaxy protoclusters yield insights into galaxy cluster formation complementary to those obtained via `archaeological' studies of present-day galaxy clusters. Submillimetre-selected galaxies (SMGs) are one class of sources used to find high-redshift protoclusters. However, due to the rarity of protoclusters (and thus the large simulation volume required) and the complexity of modeling dust emission from galaxies, the relationship between SMGs and protoclusters has not been adequately addressed in the theoretical literature. In this work, we apply the \texttt{L-GALAXIES} semi-analytic model (SAM) to the \texttt{Millennium} N-body simulation. We assign submillimetre (submm) flux densities to the model galaxies using a scaling relation from previous work, in which dust radiative transfer was performed on high-resolution galaxy zoom simulations. We find that the fraction of model galaxies that are submm-bright is higher in protocluster cores than in both protocluster `outskirts' and the field; the fractions for the latter two are similar. This excess is not driven by an enhanced starburst frequency. Instead, the primary reason is that overdense environments have a relative overdensity of high-mass halos and thus `oversample' the high-mass end of the star formation main sequence relative to less-dense environments. The fraction of SMGs that are optically bright is dependent on stellar mass and redshift but independent of environment. The fraction of galaxies for which the majority of star formation is dust-obscured is higher in protocluster cores, primarily due to the dust-obscured fraction being correlated with stellar mass. Our results can be used to guide and interpret multi-wavelength studies of galaxy populations in protoclusters.

\end{abstract}


\section{Introduction} \label{sec:intro}

Protoclusters, defined as the progenitors of present-day galaxy clusters, are powerful laboratories for studies of the assembly
of both baryonic and dark matter \citep{rover}. Studying protoclusters can yield insight into galaxy cluster formation
complementary to that obtained by studying mature clusters. For example, one can directly investigate the origin of the
intracluster medium (ICM).
Protoclusters are ensembles of dark matter halos occupying large comoving volumes at early
epochs that will eventually collapse into a galaxy cluster \citep{chiang13, lovell18}.
In simulations, in which the full history of any given dark matter halo is known, identifying the halos that will end up in a cluster by $z = 0$ -- i.e. the protocluster members -- is straightforward. In contrast, observationally identifying protoclusters is very challenging: due to the great diversity of dark matter halos' growth histories, a dark matter overdensity at high redshift may continue to grow rapidly and end up as a
massive galaxy cluster at $z = 0$, but it could also stall and wind up as an isolated massive elliptical \citep[e.g.][]{chiang13}. 

A wide variety of galaxy populations reside in protoclusters, including H$\alpha$ emitters \citep[HAEs; e.g.][]{shimakawa18a,shimakawa18b, zheng21}, Lyman $\alpha$ emitters \citep[LAEs; e.g.][]{venemans07, chiang15, jiang18, harikane19}, [OII] emitters \citep[e.g.][]{tadaki12, laishram24}, Lyman Break Galaxies \citep[LBGs; e.g.][]{toshikawa16, toshikawa18}, Balmer Break Galaxies  \citep[BBGs; e.g.][]{shi19a, shi20}, radio-loud \citep{over06, wylezalek13, hatch14, chapman23} and radio-quiet active galactic nuclei \citep[AGN; e.g.][]{boris, onoue18, stott20}, and submillimetre-selected galaxies \citep[SMGs; e.g.][]{chapman09, casey16, miller18, oteo18, zavala19}.
Protoclusters can thus be used to study diverse galaxy populations in order to understand the impact of the environment on galaxy evolution -- in particular, how environment can affect gas fueling, star formation and quenching, eventually producing populations characteristic of local clusters (see \citealt{alberts22} for a review). 

Protoclusters are typically found by searching for overdensities in galaxy surveys. Various classes of galaxies
can be used as tracers, with some being more effective than others.
However, a given structure may be overdense in terms of one galaxy population but not others.
\citet{rotermund21} found just 4 LBGs in the SPT2349-56 protocluster core at $z \sim 4.3$, which harbors more than 14 SMGs; this region would not be identified as a massive protocluster with the LBG dropout technique alone. \citet{zhang22} noted that SMGs tend to reside at the outskirts of HAE density peaks in two $z \sim 2.2$ protoclusters, and this is also the case for the Spiderweb protocluster \citep{perez-martinez23}. This phenomenon is usually attributed to assembly bias, i.e. the properties of galaxies in dark matter halos of similar mass also depend on their formation history. 
In searching for protoclusters, it is often assumed that most of the constituent galaxies will be star-forming, as is predicted by theoretical models \citep{chiang17, muldrew18, trebitsch21, fukushima22, rennehan2024}. However, recent work has revealed
concentrations of quiescent galaxies already in place by $z \sim 2.7$ \citep{mcconachie22, ito23}.

How the selection of overdense regions depends on the tracer used remains unclear. For example, there is still debate about whether AGN \citep{champagne18, uchiyama18, vicentin21} or SMGs \citep[e.g.][]{miller15,calvi23} are good protocluster tracers.
Since bright SMGs inhabit some of the most massive dark matter halos at a given epoch \citep[e.g.][]{marrone2018,Vergara2020,stach21},
one might expect that searching for overdensities of SMGs should be an effective means for finding galaxy protoclusters.
However, the rarity of SMGs ($N \sim 10^{-5}$ to $10^{-6}$ cMpc$^{-3}$ at $z \sim 2-3$; \citealt{miettinen17,dud2020}) may cause them to be incomplete tracers of protoclusters,
and `downsizing' (more massive galaxies form their stellar mass quickly and quench earlier) may result in lower-redshift protoclusters not being traced by SMGs \citep{miller15,doug20}.

To better understand how different galaxy populations trace protoclusters, one can employ mock catalogs
generated using theoretical models (i.e.~hydrodynamical simulations or semi-analytic models, hereafter SAMs).
To do this, it is necessary to select tracer populations analogous to those used to search for protoclusters observationally. This is done by predicting the spectral energy distributions (SEDs) of the model galaxies, which is challenging.
Ideally, one would simulate protoclusters
at high resolution in order to accurately capture the structure of the interstellar medium (ISM) and then perform
dust radiative transfer to compute the SEDs of the simulated galaxies.
However, because protoclusters are rare, large simulation volumes
are required, making high-resolution hydrodynamical simulations an infeasible approach. Zoom simulations of
protoclusters selected from a large N-body simulation are an alternative approach \citep[e.g.][]{bahe2017,barnes2017,bassini2020,trebitsch2021,fukushima2023,nelson2024}, but due to the computational expense of simulating extremely high-mass halos, one must either use
relatively coarse resolution  or simulate a small number of protoclusters (see \citealt{rennehan2024} for discussion). \\
\indent SAMs provide an alternative, inexpensive means to `populate' large N-body simulations with galaxies. However, for studies of observability, it is still necessary to compute SEDs for the model galaxies. In SAMs, the 3D geometry of
the model galaxies is either ignored entirely or treated in a highly idealised fashion, making directly performing radiative
transfer infeasible. However, alternative approaches can be applied. One approach is to use empirically derived SED
templates \citep[e.g.][]{somerville12}.
Alternatively, given the star formation histories (SFHs) and metallicities of model galaxies, one can perform stellar population
synthesis to compute the intrinsic SED and then attenuate the SED using a
simple model for dust attenuation.
To compute thermal dust emission, one can apply scaling relations derived by performing dust radiative transfer
on high-resolution galaxy simulations that enable accurate and efficient submm flux density predictions given a small
number of integrated galaxy properties, most importantly star formation rate (SFR) and dust mass
\citep{hayward11, hayward13, lovell21, cochrane23}. This approach can be applied to both SAMs \citep[e.g.][]{safarzadeh17}
and coarse-resolution large-volume cosmological simulations \citep{hayward21}.
By applying these well-calibrated scaling relations to model galaxy catalogs yielded by running a SAM on a large
N-body simulation, we can generate a sufficiently large mock catalog that can be used to
study and refine multiwavelength strategies to find protoclusters and understand where (environment)
and when (redshift) different galaxy populations occur. 

In this work, the first of a series, we construct a 36 deg$^2$ mock galaxy catalog by applying the \citet{henriques15}
version of the \texttt{L-GALAXIES} SAM to the \texttt{Millennium} \citep{springel05} simulation. 
We apply scaling relations from previous work to predict the submm flux density and use the resulting mock catalog to investigate the relationship between SMGs and protoclusters in the model.

The remainder of this work is
organised as follows: in Section \ref{sec:data}, we briefly describe the N-body simulation and SAM used and
how we generate the mock catalog, including optical magnitudes and submm flux densities for the model galaxies.
In Section \ref{sec:structs}, we identify the structures in the mock and define the selection of (proto)cluster members
and environments. We then investigate how SMGs trace protoclusters in the model across cosmic time in
Section \ref{sec:frac-smgs}. We investigate the physical drivers of the identified trends in Section \ref{sec:main-props}.
In Section \ref{sec:optical-counterpart}, we explore the optical properties of the model SMGs, and
in Section \ref{sec:dust-frac}, we quantify the obscured fraction
in protocluster regions and as a function of stellar mass and redshift.
We discuss our results in Section \ref{sec:discussion} and summarize our findings in Section \ref{sec:summary}. 

Throughout this work, for consistency with the cosmologically re-scaled version of the \texttt{Millennium} simulation \citep{Angulo&Hilbert15}, we assume a \emph{Planck-1} cosmology: $h = 0.673$, $\Omega_m=0.315$ and $\Omega_{\Lambda}= 0.685$ \citep{planck1}.

\section{Methodology to create the mock catalog} \label{sec:data}

We use a mock catalog constructed generally following the methodology presented in \citet{yo21}, but
with some differences, mainly in terms of the approach for predicting the SEDs of the model galaxies.
In this section, we briefly describe the main aspects of the SAM and detail the
improvements compared to the original version.

\subsection{Semi-analytic galaxy formation model}

We use the \citet{henriques15} version of the \texttt{L-GALAXIES} SAM run on the \texttt{Millennium} simulation \citep{springel05} scaled to the \emph{Planck-1} cosmology with the \citet{angulo10} algorithm. This version of \texttt{L-GALAXIES} includes a wide range of important galaxy evolution processes, such as gas infall and cooling, star formation, galaxy mergers, metal enrichment,
satellite quenching,
supermassive black hole growth, and supernova and AGN feedback (see the Supplementary Material in \citealt{henriques15}).
Our set-up produces $\sim 4$ million galaxies with $M_{\star} > 10^9 \msun$ at $z=0$ in a $(480\textnormal{ cMpc/h})^{3}$ box, from which mock galaxy samples can be extracted within pre-defined lightcones. 

In \citet{yo21}, several $\pi \ $deg$^2$ lightcones were created by configuring the line-of-sight to pass through
pre-selected structures at desired redshifts. This is helpful for studies of individual detections, as implemented
in \citet{vicentin21}. Here, we analyze a single larger lightcone of area 36 deg$^2$ extending from $z = 0-5$. The large
field of view studied here is
crucial to minimize the effects of cosmic variance and increase the number of structures within the lightcone.
However, the larger the field of view, the higher the possibility of encountering the same galaxy in the mock (replication).
In principle, this is not a problem as long as the galaxies are at different redshifts. Here, the volume corresponding to
an area of 36 deg$^2$ located between redshifts $z \sim 4.28$ and $z \sim 5.03$ (the redshifts of the last
two snapshots used in this work) is comparable to the \texttt{Millennium} volume. Extending the mock to higher
redshift would thus lead to many galaxy replications and is hence not advisable. Consequently, we do not study
higher redshifts in this work. As we will discuss in Section \ref{sec:structs}, the mock contains a sufficient number
of (proto)clusters in our redshift range of interest.

\subsection{Updated approach to predict SEDs of the model galaxies}

The previous mock version used the single-age stellar population (SSP) SED templates from \citet{maraston05},
assuming a \citet{chabrier03} initial mass function. 
We have changed our base SSP templates to the upgraded version (2016) of \citet{bruzual03}\footnote{\url{http://www.bruzual.org/\~gbruzual/bc03/Updated\_version\_2016/}}.\footnote{\citep{Lu2024} found that a stronger TP-AGB contribution was needed to reproduce the rest-frame NIR spectra of three massive quiescent galaxies at $z \sim 1-2$, suggesting that the SSP models used in this work may underpredict the contribution of TP-AGB stars. We have checked that our primary conclusions are insensitive to this issue, as the TP-AGB treatment does not significantly affect the submm flux densities of our model galaxies.} The main difference between these two is the treatment of TP-AGB stars; predicted synthetic observables considered here are similar for the two sets of SSP templates. These templates comprise 7$\times$220 SEDs, with seven metallicities from $\log{(Z/Z_{\odot})}=-2.30$ to 0.70 and 220 ages (from 0.1 Myr to 20 Gyr). The median SED wavelength resolution is $\Delta \lambda = 0.89$ \AA, which is $\sim 20$ times better than that of \citet{maraston05}.

An important \texttt{L-GALAXIES} output is the SFH arrays \citep{shamshiri15}. These
have 20 age bins, with a median bin width
equivalent to a single internal time-step for the SAM (20 times smaller than the
time interval between snapshots; see Figure 2 of \citealt{henriques20}, which shows the binning as a function of the
snapshot).
Each SFH bin contains a composite of stellar populations, which for simplicity are assumed to share the average
metallicity and formation time.
At short lookback times, the SFH bin resolution is fairly high ($\sim$ 20 Myr), but at large lookback times, this increases
to up to $\sim$ 2 $-$ 3 Gyr \citep[see][]{yates13}. 
At low redshifts, the youngest median age is $\sim 10$ Myr.
If we were to use a minimum stellar age of $10$ Myr, we would underestimate the luminosity at short wavelengths
(in particular, $\lambda < 3000$ \AA). Since the bluest wavelengths are most efficiently absorbed by dust, it is crucial to address this for robust submm predictions.

The easiest way to address this issue is to generate younger SFH bins for both the disk and bulge galaxy components in post-processing.
We call this process `SFH refinement'. In principle, from the \texttt{L-GALAXIES} SFHs, we know the difference in stellar mass between two age bins. In particular, from the youngest SFH bin, we can derive the instantaneous star formation rate, SFR($t=0$). Then, assuming this is constant within the time interval of the youngest bin, we can construct five younger SFH bins. 

\begin{figure}[ht!]
    \centering
    \includegraphics[width=\columnwidth]{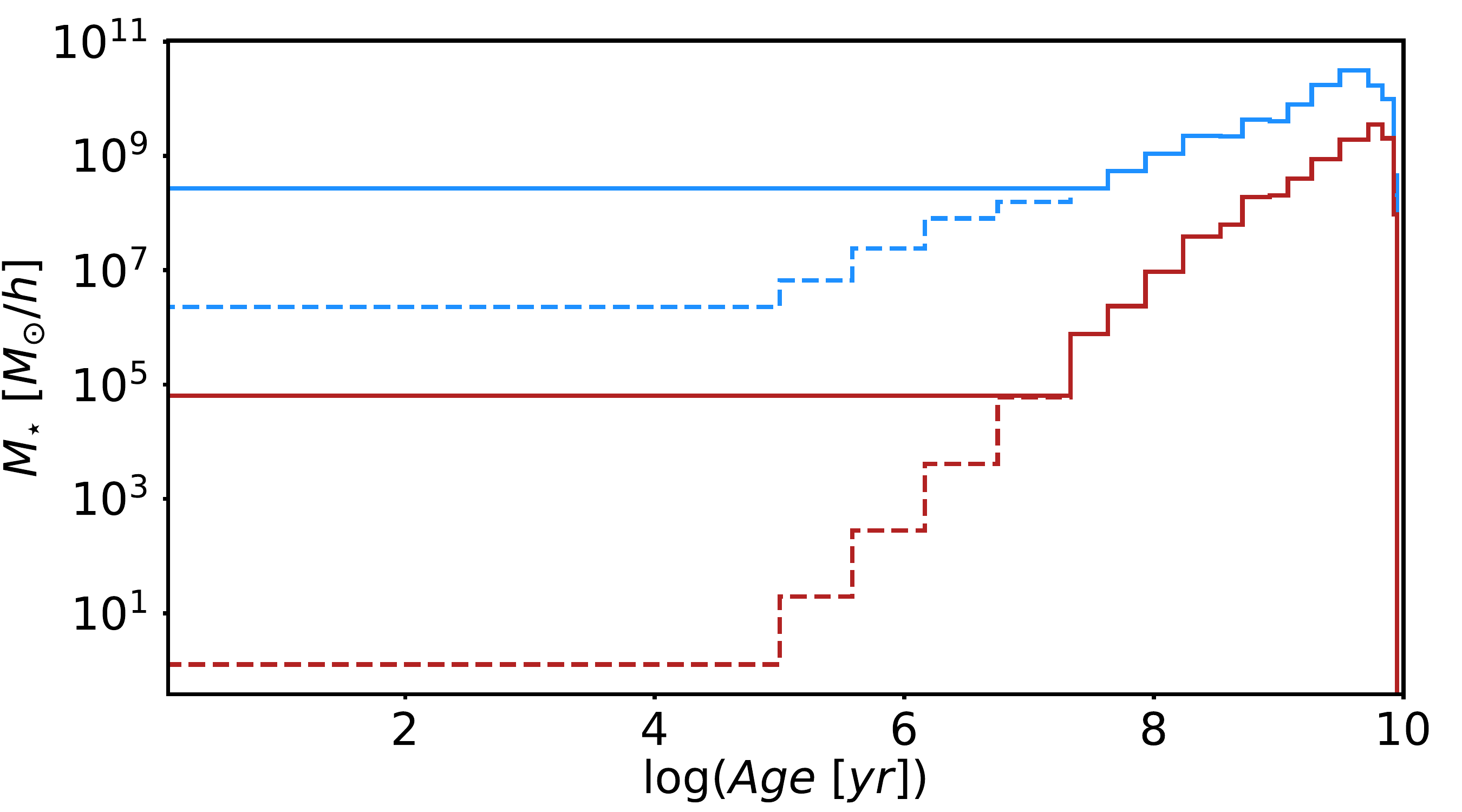}
    \caption{The stellar mass formed between two epochs with (\emph{dashed line}) and without (\emph{solid}) the SFH refinement for the disk (\emph{blue}) and bulge (\emph{red}) components. This example shows the SFH of a $z=0.4$ galaxy with $M_{\star} = 1.6\times10^{11}$ M$_{\odot}$ and SFR = 34.23 M$_{\odot}$ yr$^{-1}$. Our SFH refinement technique splits the stellar mass in the youngest (age $\sim$ 20 Myr for this snapshot) SFH bin into five bins with younger ages.}
    \label{fig:sfh_ref}
\end{figure}

In practice, we assume that we can describe SFR$(t)$ as a constant equal to SFR($t=0$) between $t=0$ and $t = t_{y,l}$, where $t_{y,l}$ is the age limit of the initial youngest SFH bin, and SFR($t = 0$) is approximated by  $M_{\star,y}/(\Delta t_{y} \  s_{\star,y})$, where $M_{\star,y}$ is the stellar mass in the original youngest SFH bin and $s_{\star,y}$ is the fraction of survivor mass given the initial SFH bin metallicity and age. We obtain $s_{\star,y}$ from the simple stellar population model used in this work.  
Then, we create five new logarithmically spaced SFH bins from 0.1 Myr to $t_{y,l}$, and ascribe to each new bin a stellar mass obtained by simply multiplying SFR($t=0$) to the time width of each new steps. Notice that the new SFH bins represent the \emph{total} stellar mass. Then, we re-correct them to obtain the survivor mass at the new SFH ages. Figure \ref{fig:sfh_ref} shows the new SFH bins produced with our method for the disk and bulge components of an example galaxy (blue and red dashed lines, respectively), alongside the original bins (solid lines).  
We assume that all of these new bins have the same metallicity, equal to the cold gas metallicity. 

Another upgrade to the method to derive galaxy SEDs is the implementation of intergalactic medium absorption. Instead of using pre-computed tables, i.e. magnitude corrections as a function of redshift, we have applied a transmission curve to each modeled galaxy SED. To do this, we constructed a grid with the IGM transmission curves for sources at redshifts from $z=0.1$ to $z=7.0$ (although our mock goes up to $z \sim 5$) spaced by $\Delta z = 0.1$. We used the \texttt{IGMTransmission} code \citep{harrison11} to obtain standardized curve sets. This algorithm performs Monte Carlo simulations to distribute absorbers along a line-of-sight, where we sample their redshift and optical depth following the \citet{inoue08} model. This component is crucial for obtaining reliable magnitudes at optical wavelengths for high-z galaxies, since features such as the Lyman break and Lyman-$\alpha$ forest are detected in the $u$ and $g$ bands, respectively, for $z \sim 3.0$ sources.

\subsection{Dust attenuation model} \label{sec:atten_model}

Because it is particularly relevant for this work, we describe the dust attenuation model employed,
based on that of \citet{charlot00}, in detail here.
The same model has been employed in numerous previous works \cite[e.g.][]{henriques15,shamshiri15, clay15,henriques20,yo21}.
The dust model assumes two components: the ISM and molecular clouds (MCs; also called birth clouds)
around recently formed stars. The ISM attenuation affects the entire stellar population of the disk, whereas
the MC attenuation acts only on the light from young stars (age $\leq 10$ Myr). 

The optical depth as a function of the wavelength for the ISM component is
\begin{equation}
\tau_{\lambda}^{\rm ISM} = \left ( \frac{A_{\lambda}}{A_V} \right )_{Z_{\odot}}
\left ( \frac{Z_{\rm gas}}{Z_{\odot}} \right )^s (1 + z)^{-1}
\left ( \frac{\langle N_H\rangle}{2.1 \times 10^{21} \ \mathrm{cm}^{-2}} 
\right ).
\label{tau_ism}
\end{equation}
The mean hydrogen column density, $\langle N_H \rangle$, can be obtained directly from
the SAM output parameters as follows:
\begin{equation}
\small
\langle N_H\rangle = \frac{M_{\rm cold}}{1.4 m_p \pi (a R_{{\rm gas, d}})^2} \
\mathrm{cm}^{-2} ,
\label{n_h}
\end{equation}
where $M_{\rm cold}$ is the mass of the cold gas, $R_{{\rm gas, d}}$ is the radius of the gaseous disk, $m_p$ is the proton mass,
and $a = 1.68$.
This value for $a$ is selected so that $\langle N_H \rangle$ represents the mass-weighted average column density
of an exponential disk. The factor $1.4$ takes into account the helium abundance \citep{clay15}. The $(Z_{\rm gas}/Z_{\odot})$ factor in Equation \ref{tau_ism} is the mass fraction of metals in the cold gas in solar metallicity units (we use $Z_{\odot} =$ 0.02 for consistency with \citet{henriques15}). In this model, the index for the power-law dependence on metallicity, $s$,
depends on the wavelength: $s=1.35$ for $\lambda < 2000$ \AA , and $s=1.60$ for $\lambda \geq$ 2000 \AA~ \citep{guiderdoni87}.
$(A_{\lambda}/A_V)_{Z_{\odot}}$ represents the solar-metallicity extinction curve from \cite{mathis83}.

The optical depth of the MC component is
\begin{equation}
\tau_{\lambda}^{\rm MC} = \tau_V^{\rm ISM} \left ( \frac{1}{\mu} - 1 \right ) \left ( \frac{\lambda}{5500 {\rm \AA}} \right )^{-0.7} ,
\label{tau_bc}
\end{equation}
where $\tau_V^{\rm ISM}$ is the optical depth of the ISM in the $V$ band ($ \lambda_{\rm eq} \sim 5500$  \AA ), and $\mu$ is a random Gaussian variable with values between $0.1$ and $1$, with mean $0.3$ and standard deviation $0.2$ \citep{charlot00}.

Therefore, the attenuation as a function of wavelength is given by
\begin{equation}
A_{\lambda}^{\mathrm{ISM}} = \left ( \frac{1 - e^{(-\tau_{\lambda}^{\mathrm{ISM}}\sec{\theta})}}{ \tau_{\lambda}^{\mathrm{ISM}}\sec{\theta}} \right ); \ 
A_{\lambda}^{\mathrm{MC}} = (1 - e^{-\tau_{\lambda}^{\mathrm{MC}}}) ,
\label{dusts}
\end{equation}
where $\theta$ represents the inclination of the galaxy. The inclination cosine is first randomly sampled between $0$ and $1$.
Then, all values less than $0.2$ are set to $0.2$ \citep{henriques15}. 

\subsection{Dust emission modeling} \label{sec:dust_emission}

The thermal dust emission SED of a galaxy is determined by the radiation field heating the dust, dust density field, and grain properties. If the
3D stellar and AGN radiation field, 3D dust density field, and grain properties are known, one can perform radiative transfer to compute the dust
temperature distribution and thus dust emission SED. In a SAM, one can assume an idealised geometry and perform radiative transfer
\citep[e.g.][]{silva1998}. However, to avoid introducing additional free parameters, we adopt an alternative approach: we use scaling relations derived from performing radiative transfer on hydrodynamical simulations.
We present a brief comparison with an empirical SED template-based approach to demonstrate that our method yields reasonable results.

\subsubsection{Scaling relations} \label{subsec:scaling_relation}

The dust temperature distribution, and thus dust emission SED, is affected by the total luminosity absorbed by dust and the dust mass.
Multiple authors have performed radiative transfer in post-processing on galaxies from simulations that differ considerably in terms of
spatial resolution, the galaxy formation model employed, and the code used and found that the submm flux density can be predicted
reasonably well if only the total infrared (IR) luminosity and dust mass are known \citep{hayward11, hayward13, lovell21, cochrane23}.
Parameterising submm flux density as a double power law in SFR (i.e.\,assuming AGN heating of dust is subdominant) and dust mass, 
they found similar values for the free parameters, suggesting that although simple, this approach is a robust method for predicting submm
flux density in models.

Here, we will use the fit derived by \citet{cochrane23}, whose free parameters were obtained by applying the Stellar Kinematics Including Radiative Transfer (\texttt{SKIRT}) Monte Carlo radiative transfer code \citep{baes11, camps15} to massive, high-redshift galaxies drawn from the Feedback in Realistic Environments
(FIRE) project (\citealt{hopkins14, hopkins18}; see also \citealt{cochrane19}). In addition to SFR and dust mass, the relation from \citet{cochrane23} uses stellar mass and redshift as input
quantities to predict the observed-frame 870-$\mu$m flux density. Their relation is the following:
\begin{equation} \label{cochrane_relation}
\small
    \frac{S_{870}}{\rm mJy} = \alpha \left (  \frac{\rm SFR}{100 \msun {\rm yr}^{-1}}\right )^{\beta} \left (  \frac{M_{\star}}{10^{10} \msun}\right )^{\gamma} \left (  \frac{M_{\rm dust}}{10^{8} \msun}\right )^{\delta} (1 + z)^{\eta},
\end{equation}
where $\log \alpha = -0.77$, $\beta = 0.32$, $\gamma=0.13$, $\delta=0.65$, and $\eta=0.65$. 
The version of \texttt{L-GALAXIES} used here does not provide dust mass as an output.
Thus, we simply assume that 40 percent of the metals in the cold gas are in the form of dust grains, $M_{\rm dust} = 0.4\,M_{\rm Z,ColdGas}$ \citep{dwek98},
which is a reasonable approximation for the relatively metal-enriched, massive galaxies studied in this work, as seen in observations \citep[e.g.][]{remyruyer14,choban22}, and in later versions of \texttt{L-GALAXIES} which incorporate dust physics \citep{Vijayan+19,Yates+24}.

\citet{cochrane23} show that their scaling relation can recover the true values of submm flux density from the radiative transfer calculations with an error of
$\sim 0.1$ dex, independent of flux density. To incorporate this error, when assigning $S_{870}$ values to the model galaxies using Equation \ref{cochrane_relation},
we add values drawn from a normal distribution with zero mean and standard deviation of 0.1 dex.

\begin{figure}
    \centering
    \includegraphics[width=\columnwidth]{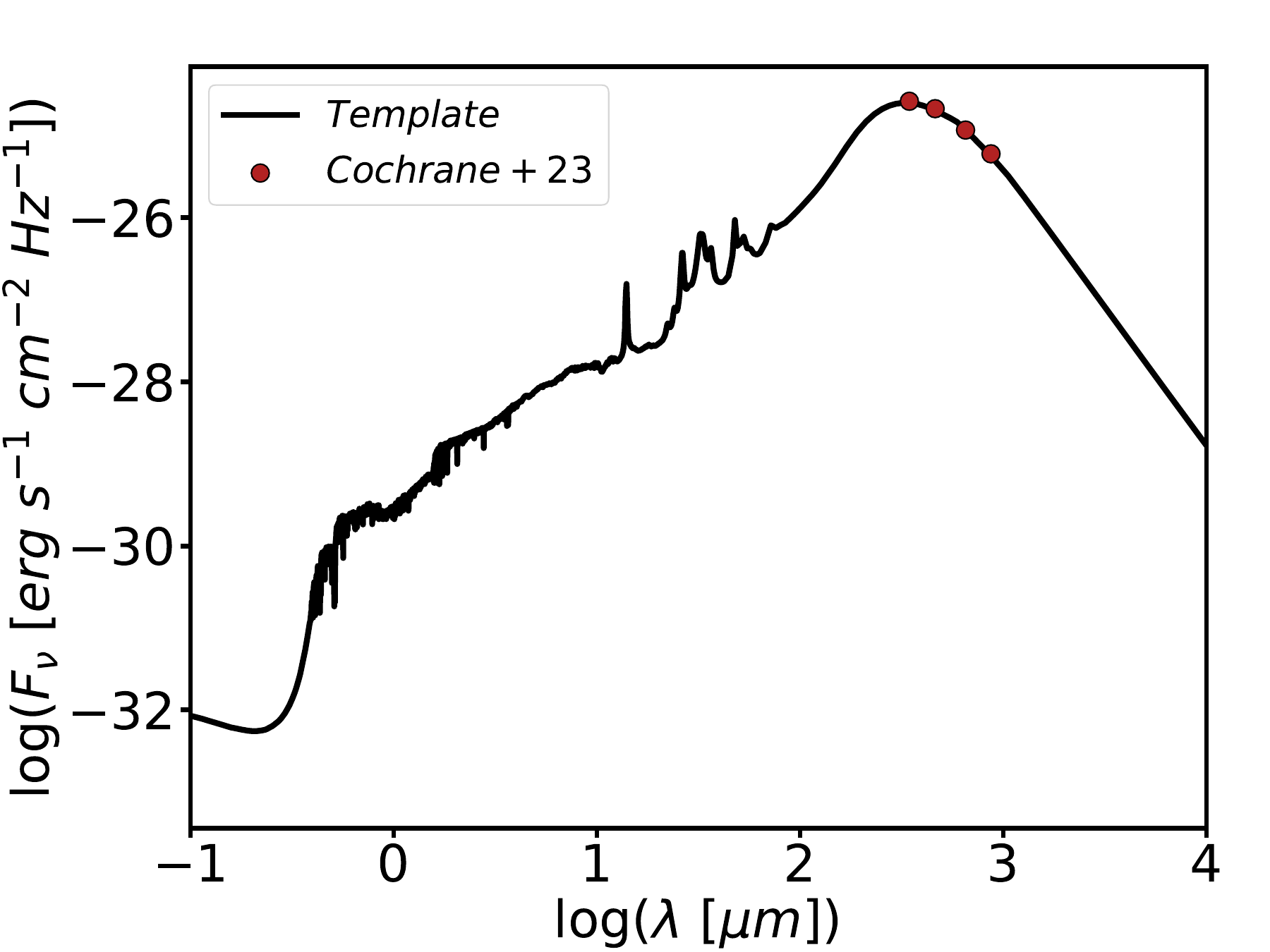}
    \caption{Example SED of a galaxy at $z=3.23$ with $\log_{10} (M_{\star}/\mathrm{M_{\odot}}) = 11.14$, ${\rm SFR} = 400.27 \msun ~{\rm yr}^{-1}$, and $\log_{10}(M_{\rm dust}/\mathrm{M_{\odot}}) = 9.19$ computed using the template approach. Red dots indicate the flux densities computed using the \citet{cochrane23} scaling relations.
    The template and scaling relation approaches yield similar results.  
    \label{fig:sed_example}}
\end{figure}

\subsubsection{Template approach} \label{sec:templates_app}

An alternative method for modelling the thermal dust emission SED is using templates.
Although we use the scaling relation approach as our fiducial method, we also derive submm flux densities using empirical templates to determine
the sensitivity of our results to how submm flux density is computed. We use the \citet{boquien21} templates, which are parameterised by the
specific SFR (sSFR $\equiv$ SFR$/M_{\ast}$) and the total IR luminosity of the galaxy. We obtain the latter by integrating the total luminosity absorbed by dust (see Section \ref{sec:atten_model}).

For illustration, we present in Figure \ref{fig:sed_example} the full SED of a galaxy at $z=3.23$ computed using the template approach and overplot flux densities at a number of rest-frame far-IR wavelengths using the \citet{cochrane23} scaling relations. We show flux densities in ALMA bands 10, 9, 8, and 7
(observed-frame effective wavelengths of 345 $\mu$m, 426 $\mu$m, 652 $\mu$m, and 870 $\mu$m, respectively) because
\citet{cochrane23} derived scaling relations for each of these bands. The two approaches yield very similar results.

\begin{figure}[ht!]
\includegraphics[width=\columnwidth]{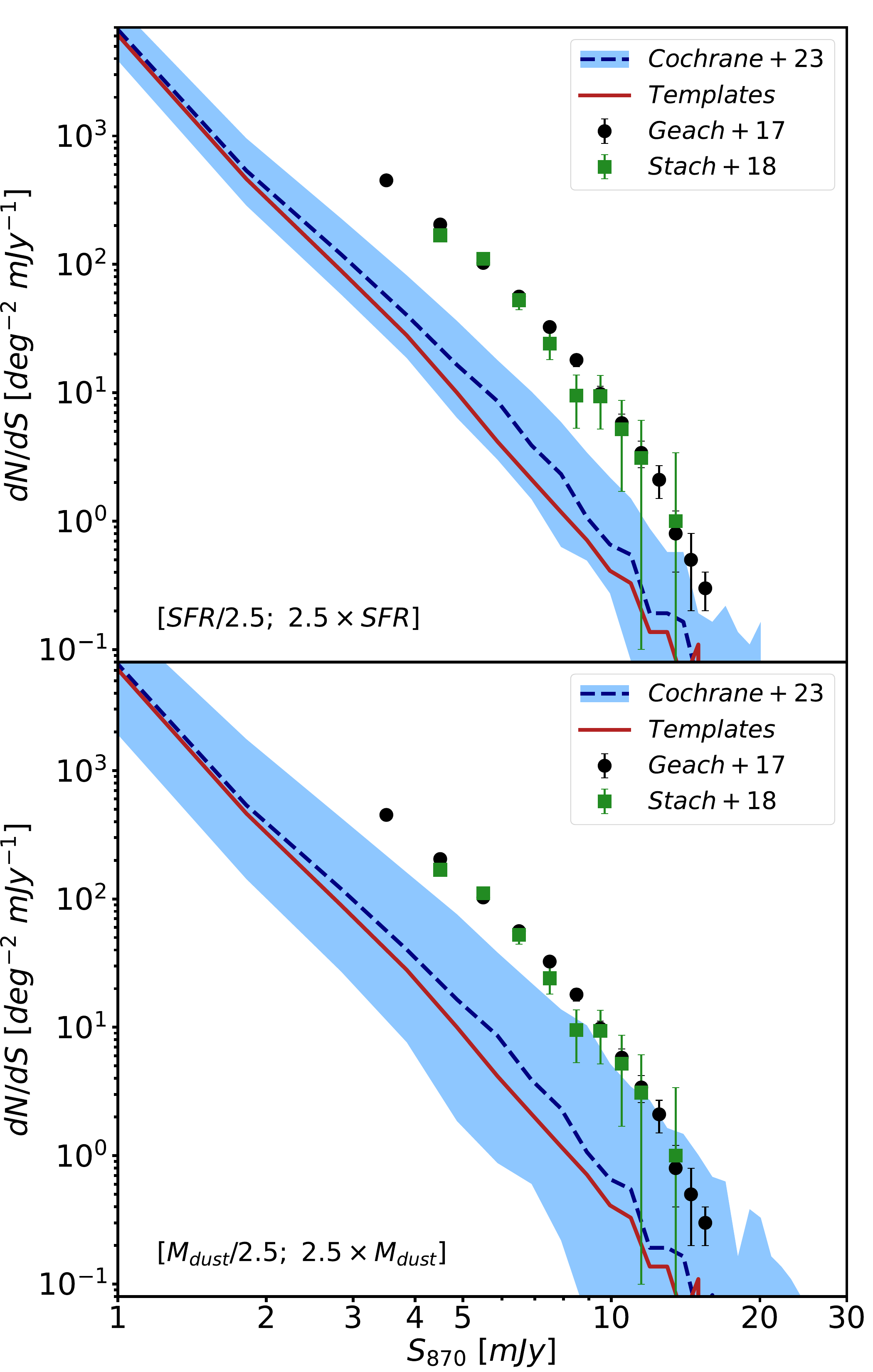}
\caption{The predicted $S_{870}$ differential number counts for our \texttt{Millennium}+\texttt{L-GALAXIES} mock, computing submm flux densities using both the \cite{cochrane23} scaling relation (\emph{blue dashed line}) and the template approach (\emph{red solid line}). We overplot observational results from \cite{geach17} and \cite{stach18}. For reference, colored areas represent the predicted number counts from the \cite{cochrane23} scaling relation if we arbitrarily divide or multiply the SFR (\emph{top panel}) or $M_{\rm dust}$ (\emph{bottom panel}) by a factor of 2.5. Note that multiplying $M_{\rm dust}$ by 2.5 implies that all metals in the cold gas would be dust. The two approaches for computing submm flux density yield similar number counts. As is the case with most
models that assume a standard IMF, the predicted counts are less than those observed, though only by a factor of a few.
\label{fig:submm_ncounts}}
\end{figure}

\subsubsection{Comparison with observations} \label{sec:obs_comp}

In Figure \ref{fig:submm_ncounts}, we plot the differential submm number counts (i.e.\,integrated over redshift) predicted using the two approaches for computing $S_{870}$ and
compare the predictions with observations from \citet{geach17} and \citet{stach18}.

The two methods for computing $S_{870}$ yield similar predictions for the submm number counts (the lines in Figure \ref{fig:submm_ncounts}).
As is the case for most models that assume a standard IMF \citep[e.g.][]{hayward13,hayward21,lovell21},
we underpredict the observed counts. To explore the severity of this discrepancy, we show how much the differential counts change when
we artificially increase or decrease the SFR (top panel) or dust mass (bottom panel) by a factor of 2.5.
In the case of the SFR, our motivation for this factor comes from the fact that the cosmic star formation density in
the \citet{henriques15} version of the \texttt{L-GALAXIES} model is less than the empirically derived values from \citet{behroozi13} and \citet{madau14}
by a similar factor.
For the dust mass, since we have implemented a constant cold gas dust-to-metals ratio equal to 0.4,
the factor of 2.5 would represent the extreme case in which all metals in the cold gas phase are locked up in dust.
We can see that boosting the SFR and dust mass simultaneously could lead to number counts in agreement with those observed. In follow-up work,
we will examine what modifications to the model are necessary to match the observed submm number counts while maintaining agreement
with the other observational constraints employed. In the present work, we simply use the version of \texttt{L-GALAXIES} from
 \citet{henriques15} and focus on differential quantities, such as the enhancement in the submm-bright fraction in protocluster cores
 compared with the field, which are unlikely to be affected by the overall underprediction of the submm counts.

\section{Identifying (proto)clusters in the mock} \label{sec:structs}

\begin{figure}[ht!]
\includegraphics[width=\columnwidth]{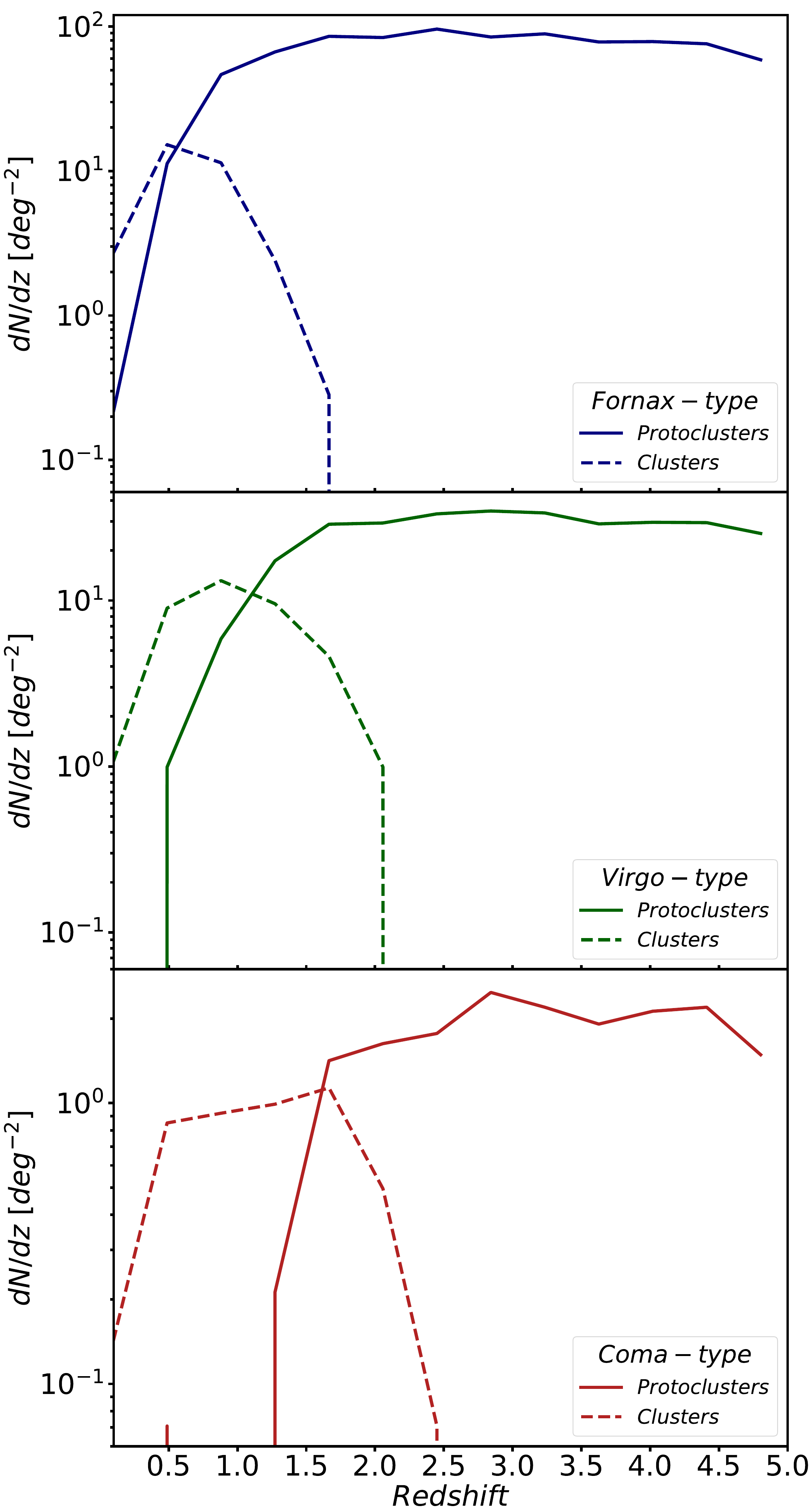}
\caption{The number counts (deg$^{-2}$) of protoclusters (\emph{solid lines}) and clusters (\emph{dashed lines}) as a function of redshift.
Once a protocluster reaches a virial mass of $10^{14}\msun$, it is subsequently classified as a cluster.
Fornax-, Virgo-, and Coma-type progenitors are shown in the \emph{top, middle, and bottom panels}, respectively.
The redshift at which the
numbers of protoclusters and clusters are the same is $z \sim 0.7$, 1.2, and 1.7 for Fornax-, Virgo-, and Coma-type
progenitors, respectively.
\label{fig:struct_zdist}}
\end{figure}

\begin{figure*}[ht!]
\includegraphics[width=\textwidth]{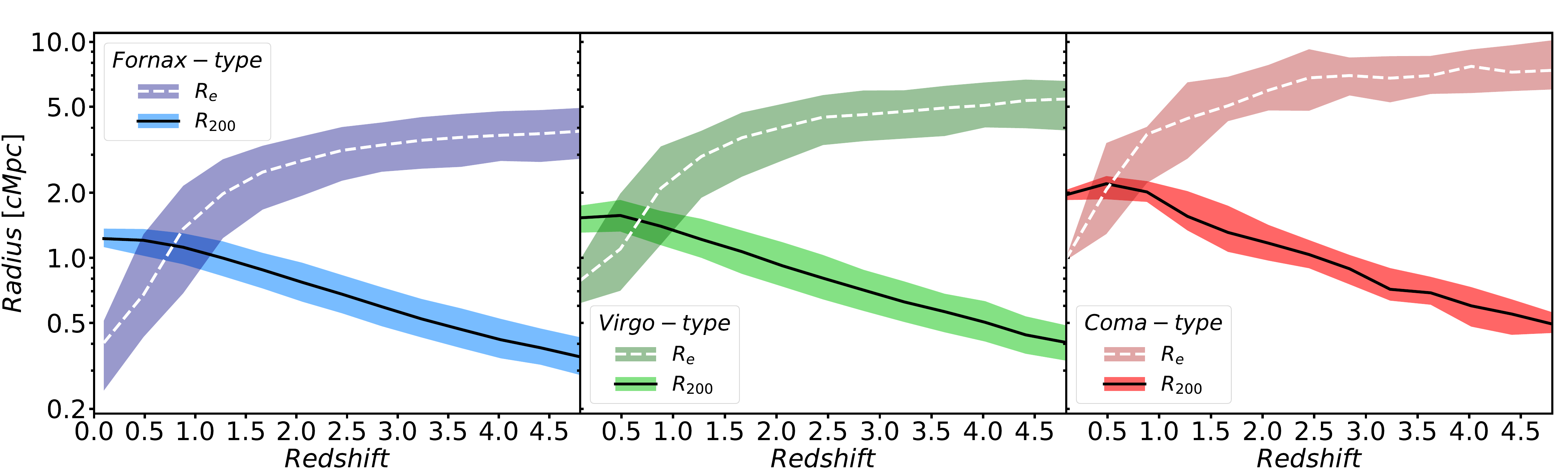}
\caption{The evolution of the effective radius ($R_e$) and the most massive dark matter halo's radius ($R_{\mathrm{200c}}$) as a function of redshift. We show Fornax-, Virgo- and Coma-type progenitors from left to right. Colored areas represent the 25th and 75th percentiles, respectively, while the lines are the medians. As shown in \citet{chiang13}, $R_e$ increases with increasing redshift. On the other hand, $R_{\mathrm{200c}}$ decreases with increasing redshift and becomes comparable to $R_e$ at $z \sim 1$. Thus, we will focus our analysis at $z \gtrsim 1.5$ since our galaxy member selection depends on both quantities.
\label{fig:radius_ev}}
\end{figure*}

In order to identify the predicted (proto)clusters within the lightcone, we have to link them to the dark matter halo information of the base simulation. We first define clusters following \citet{chiang13}: principal dark matter halos with \texttt{M\_tophat} $ > 1 \times 10^{14} \ \msun$. There are 3,279 such clusters at $z=0$. 

In the second step, we select all galaxies hosted by $z = 0$ cluster progenitors halos using the \texttt{haloId} key identifier. Note that we could adopt an intermediate step where we identify the other bound subhalos of those progenitor clusters (e.g. by matching their \texttt{FOFCentralID}), but we verified that this does not make any significant difference for the structures at $z > 1$.

As mentioned in Section \ref{sec:data}, galaxy replication can occur, and we can find up to six progenitors (with a median value of four) of the same galaxy in the mock but at different redshifts and celestial coordinates. Consequently, the procedure described above to identify structures in the mock by searching for cluster progenitors reflects the effect of replications. The redshift distribution of galaxies hosted in progenitor halos of a given $z = 0$ cluster exhibits peaks where replication occurs. Although replicated galaxies are progenitors of the same cluster, they are effectively different structures since they are at distinct evolutionary stages. We thus identify these peaks to separate them into single structures. To do this, we first construct redshift histograms with a bin size equal to $\Delta z= 0.01$. If the number of galaxies in a redshift bin is greater than the numbers for the two adjacent ones, we consider this redshift to be the initial structure redshift, $z_i$. To obtain accurate (proto)cluster redshifts and celestial coordinates, we select all galaxies (in progenitor halos) with redshift $z_i \pm 5 \Delta z$ and then compute the median right ascension, declination, and redshift. With this approach, we identify 16,320 structures within the 36 deg$^2$ mock at $z \lesssim 5.0$.

Here, we adopt the same classification of (proto)clusters as in \citet{chiang13}, based on their descendant cluster's $z = 0$ mass ($M_{z = 0}$). We refer to protoclusters as Fornax-, Virgo-, or Coma-type if their $z=0$ descendant cluster masses are $M_{z = 0} =$ (1.37 - 3)$\times 10^{14}$, (3 - 10)$\times 10^{14}$,
or $\geq 10^{15}$ $\msun$, respectively.\footnote{This terminology is simply a convenient way to put descendant masses in context. We do not intend to imply that Coma-type protoclusters will resemble the Coma cluster at $z = 0$ in terms of any property except for mass or that the detailed formation histories of a given protocluster type are similar.}
Additionally, following  \citet{chiang13}, a protocluster becomes a cluster once its main dark matter halo has virial mass \texttt{M\_tophat} $ > 10^{14}$ $\msun$. We show the redshift distributions of the structures in the mock in Figure \ref{fig:struct_zdist}.

In total, our sample (defined purely in terms of mass, as described above -- we do not yet consider whether galaxies are submm-bright) comprises the following: 
\begin{itemize}
    \item Fornax-type: 452 clusters; 12,080 protoclusters.
    \item Virgo-type: 542 clusters; 4,199 protoclusters.
    \item Coma-type: 65 clusters; 247 protoclusters.
\end{itemize}

Although we have divided the sample of $z=0$ cluster progenitors into already-formed clusters and protoclusters to construct Figure \ref{fig:struct_zdist}, we do not employ this separation in the remainder of our analysis, and for convenience, we will use the term `protocluster'
even though some of the structures already have mass $ > 10^{14}$ $\msun$ at high redshift.
This is because the definition of a cluster purely based on mass ignores other important characteristics of clusters, such as whether a hot intracluster
medium has formed.
In practice, this is important only at $z \la 1.5$, as above that redshift, protoclusters with \texttt{M\_tophat} $ > 10^{14}$ $\msun$ constitute a negligible
fraction of the population.

Since we know the full dark matter halo merger history in the simulation, we can unambiguously define which galaxies are protocluster
members using the \texttt{haloId}. Since this is not possible for the real universe, we test spatially based criteria
to select the protocluster galaxies considered to be members of a given protocluster. 

First, we estimate the effective radius of each structure. \citet{chiang13} defined this as the second moment of the member galaxies' positions weighted by their halo masses. Instead, for simplicity when working with a mock galaxy catalog, we define the effective radius as follows:
\begin{equation}
    R_e = \sqrt{\frac{1}{M_{\star, {\rm tot}}} \sum_i M_{\star,i}({\bf x_i} - {\bf x_c})^2 },
\end{equation}
where $M_{\star,i}$ is the stellar mass of a member galaxy located at ${\bf x_i}$, $M_{\star, {\rm tot}}$ is the sum of all $M_{\star,i}$, and ${\bf x_c}$ is the center of mass of the protocluster at its redshift. 
Another important radial scale, which we will use to define core galaxies, is the virial radius for the
most-massive progenitor halo of a given protocluster, $R_{\mathrm{200c}}$. We show the evolution of $R_e$ and $R_{\mathrm{200c}}$ as a
function of redshift for the three descendant types in Figure \ref{fig:radius_ev}. 

To investigate potential environmental effects, which should be stronger in the central, highest-density
regions of protoclusters, we distinguish protocluster cores and outskirts.
Core galaxies are all galaxies located within a sphere of radius $R_{\mathrm{200c}}$ centered at the position
of the central (most-massive) galaxy.
We define protocluster outskirts galaxies as all galaxies enclosed in a sphere centered at the structure's center of mass with a radius $2 R_e$, excluding core galaxies.\footnote{We have confirmed that our results are qualitatively insensitive to the definitions of `core' and `outskirts.'}
Figure \ref{fig:radius_ev} shows that for $z \ga 1$, $R_{\mathrm{200c}} < R_e$, so this definition of outskirts is sensible. At lower redshift, it would no longer work. Thus, we will focus our analysis at $z \gtrsim 1.5$ to avoid the difficulties of mixing core and outskirt galaxies that arises with the spatially based criterion.

\begin{figure}[ht!]
\includegraphics[width=\columnwidth]{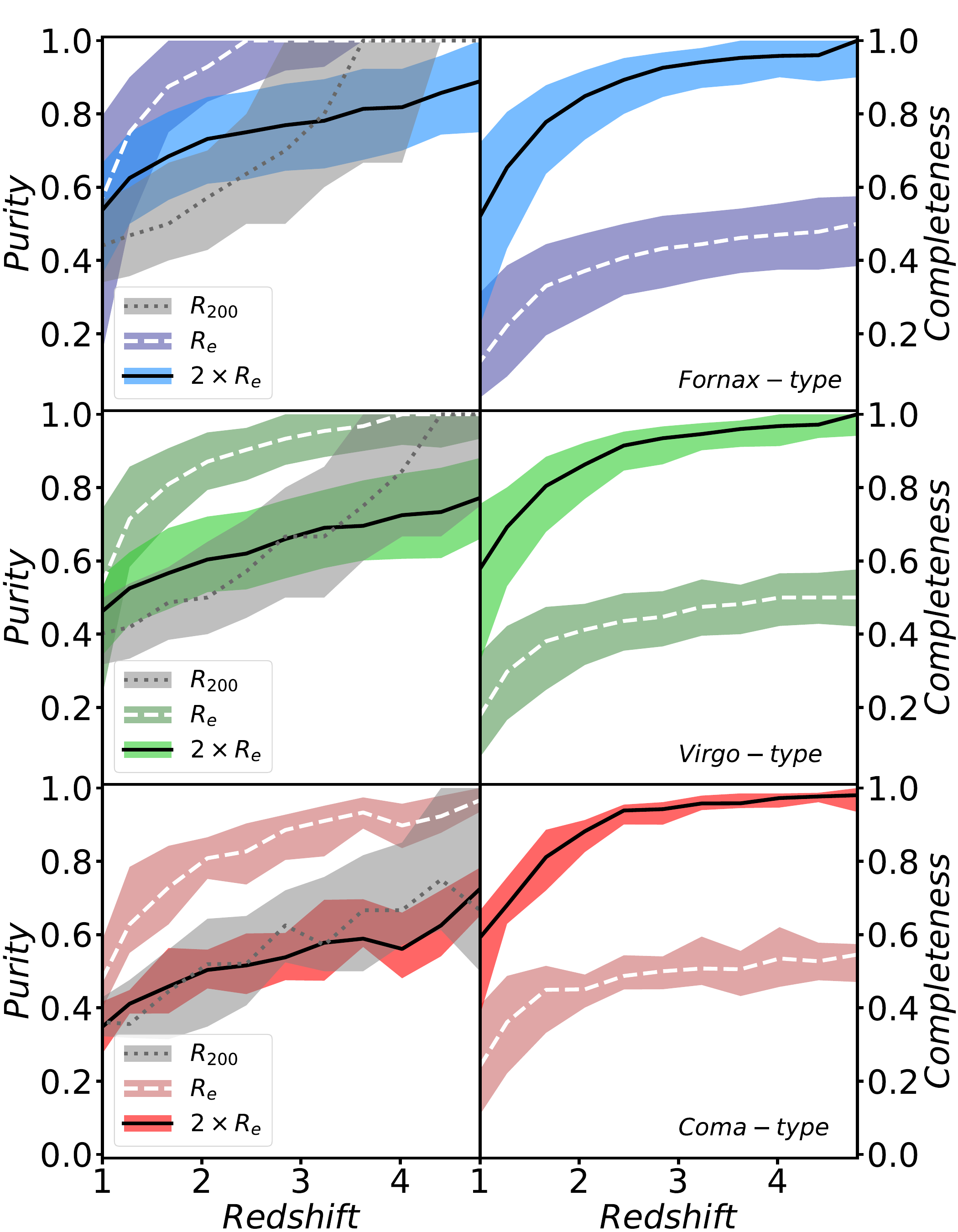}
\caption{
Purity (\emph{left column}) and completeness (\emph{right column}) as a function of redshift for the three protocluster classes (\emph{top:} Fornax-type;
\emph{middle:} Virgo-type; \emph{bottom:} Coma-type) when using the spatially based criteria to select protocluster members; whether a galaxy is
actually a protocluster member is known from its dark matter halo merger history. The lines indicate the median value for the protocluster sample, and the shaded region indicates the 25-75th-percentile range.
The results for $R_e$ and $2 R_e$ consider all protocluster members, whereas those for $R_{\mathrm{200c}}$
consider only members within the core region.
Using a radius of $2 R_e$ to select protocluster galaxies yields an overall completeness of 80 percent, while with $R_e$, this quantity is generally less than $\sim 50$ percent. The purity is $\ga 80$ percent when $R_e$ is used, whereas it is $\sim$50-80 percent when $2 R_e$ is employed, depending
on redshift and protocluster type.
When selecting core galaxies within $R_{\mathrm{200c}}$,
the purity drops with decreasing redshift, mainly due to $R_{\mathrm{200c}}$ becoming comparable with $R_e$ and thus enclosing increasingly many galaxies
that will not merge with the main halo by $z = 0$.
\label{fig:pur_com_gmem}}
\end{figure}

To test the ability of the spatially based criterion to select member galaxies,
we compute the purity and completeness of member selection for protocluster cores and outskirts as defined above.
We quantify the purity and completeness as a function of redshift for the three cluster progenitor classes. We note that the purity can be less than 100 percent for $R_{\mathrm{200c}}$ because a galaxy may be located within $R_{\mathrm{200c}}$ of the central galaxy but still remain a distinct subhalo at $z =$ 0. Such a galaxy would be considered part of the protocluster outskirts but not the core and thus be considered a contaminant when computing the purity of the $R_{\mathrm{200c}}$ selection. This effect is especially prominent at low redshift, where $R_{\mathrm{200c}}$ can exceed 1 cMpc.
We present the results in Figure \ref{fig:pur_com_gmem}.

Figure \ref{fig:pur_com_gmem} shows that as expected, when we use a larger radius ($2\times R_e$) to select the member galaxies, we
obtain higher completeness but lower purity. For this choice of radius, the purity is $\sim{}50-60$ percent at $z\sim{}1.5$, and the completeness is $\ga 60$ percent. For $R_{e}$, the purity is $\sim$60-80 percent, but
the completeness is $\la 40$ percent. At higher redshift, the completeness
for $2R_{e}$ is $\ga90$ percent, whereas for $R_e$, it is at most
$\sim50$ percent. We thus opt to use a radius equal to $2R_e$ to define
protocluster `outskirts', but we have confirmed that the results are
qualitatively the same when $R_e$ is used.

\section{Results} \label{sec:results}

In the following sections, we explore the dependence of protocluster galaxy populations on redshift and environment. 

\subsection{An excess of SMGs in protocluster cores} \label{sec:frac-smgs}

\begin{figure*}[ht!]
\includegraphics[width=\textwidth]{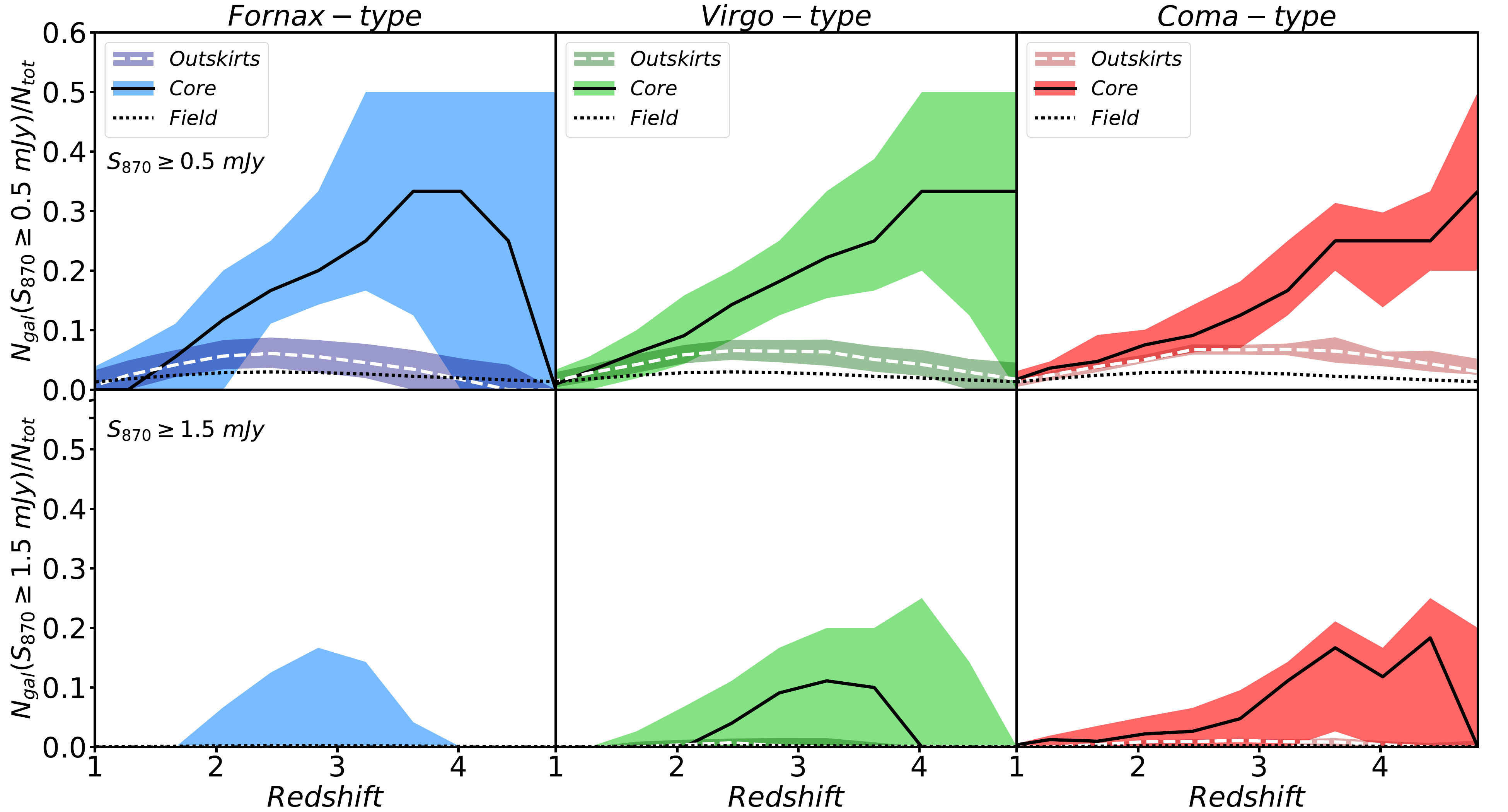}
\caption{The redshift evolution of the fraction of galaxies with $S_{870} \geq 0.5$ mJy (\emph{first row}) and $S_{870} \geq 1.5$ mJy (\emph{second row}) in protocluster cores (\emph{solid lines})
and protocluster outskirts (\emph{dashed lines}) for the three protocluster types (\emph{left:} Fornax-type; \emph{middle:}
Virgo-type; \emph{right:} Coma-type).
The lines denote the median value at a given redshift, and the shaded regions indicate the 25-75th-percentile range.
The submm-bright fraction for the field is indicated by the dotted line.
At $z \ga 1.5$, protocluster cores exhibit an enhanced submm-bright fraction relative to both protocluster outskirts and the field.
The redshift evolution of the submm-bright fraction in the core (both the peak and the duration) presents a clear dependence on the protocluster type.
Protocluster outskirts have submm-bright fractions only slightly elevated relative to the field.
For the higher flux density threshold ($S_{870} \geq 1.5$ mJy), the same trend is present, but all fractions are lower. 
\label{fig:frac_submm}}
\end{figure*}

We first consider the fraction of submm-bright galaxies in different environments to investigate the connection between SMGs
and protoclusters. We analyze mock galaxies with $S_{870}$ greater than two different flux limits, 0.5 mJy or 1.5 mJy. The lower limit is motivated by the flux density of the faintest submm-detected members of the SPT2349-56 protocluster \citep{miller18}, as measured using ALMA. 
First, for each protocluster,
we quantify the numbers of SMGs in the core and in the outskirts brighter than a given flux density. Then, we divide by the total number of galaxies in the protocluster core and outskirts to compute the submm-bright fractions for each region, $N_{\rm gal}(S_{870} \geq x)/N_{\rm tot}$, where $x$ is 0.5 mJy or 1.5 mJy. We also compute this quantity for the field (i.e.\,all galaxies in the mock in a
given redshift bin that do not reside in protoclusters).
The panels of Figure \ref{fig:frac_submm} show how the submm-bright fraction depends on redshift
for the different protocluster types. The top row shows
the results for a flux cut of $S_{870} > 0.5$ mJy, whereas the bottom shows the results for a flux density cut of 1.5 mJy.

The most notable trend in Figure \ref{fig:frac_submm} is that at $z \ga 1.5$, the submm-bright fraction is higher in protocluster
cores than in both protocluster outskirts and the field.
We have confirmed that this result is qualitatively insensitive to the definition of `cores' (e.g.\,using $R_{\mathrm{500c}}$ instead of $R_{\mathrm{200c}}$).
The redshift at which the submm-bright fraction for protocluster cores peaks is higher for more massive clusters.
The peak of $N_{\rm gal}(S_{870} \geq 0.5$ mJy$)/N_{\rm tot}$ for Coma-type protoclusters lies at a higher redshift than the limit of the mock ($z>5$). In the case of Virgo-type protoclusters, the peak submm-bright fraction is at $z \sim 4$, while for Fornax-type protoclusters, it is at $z \sim 3.5$. On the other hand, the slope of this fraction for redshifts lower than the peak is higher for low mass cluster progenitors. This implies that the redshift range of the submm phase of protocluster cores is higher for more massive systems.

Galaxies in the outskirts of protoclusters present a slightly higher fraction of SMGs compared to the field over the entire redshift range, but the value is always much closer to that of the field than it is to that for cores. As in the case of the protocluster cores, the redshift at which the fraction of SMGs in protocluster outskirts peaks is slightly higher for the progenitors of more massive clusters.

We find similar trends in the evolution of submm-bright fractions and their environmental dependence for the two different flux density limit thresholds ($S_{870}>0.5$ mJy and $S_{870}>1.5$ mJy).
Notice that the $N_{\rm gal}(S_{870} \geq 1.5$ mJy$)/N_{\rm tot}$ values for the field and protocluster outskirts are close to zero.
For both flux density cuts, the values for the submm-bright fractions would increase if our model reproduced the total submm
number counts, but we expect that the differences are robust;
see Section \ref{dis:underprediction} for further discussion.
Given the paucity of sources with $S_{870} > 1.5$ mJy, for the rest of this work, we only consider the $S_{870}>0.5$ mJy
definition of `submm-bright'.

\begin{figure*}[ht!]
\includegraphics[width=\textwidth]{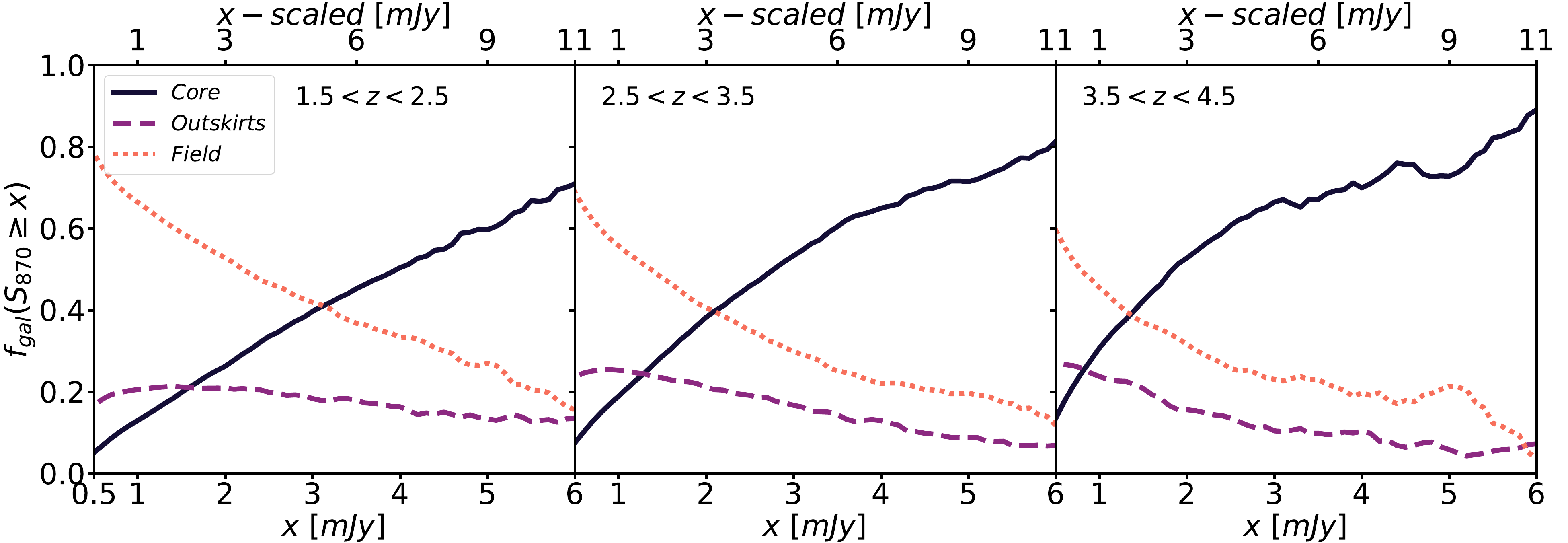}
\caption{The fraction of SMGs with $S_{870}$ higher than x mJy that are in the field (\emph{orange dotted line}), protocluster outskirts (\emph{purple dashed}), or cores (\emph{black solid}) at $z \sim 2$, 3, and 4 (from left to right). The values on the top $x$-axes are rescaled flux densities, where we have boosted all model galaxies' flux densities by a factor of 1.81; see text for details. At the lowest flux densities
considered, most SMGs are located in the field. Above a transition flux density, which decreases with increasing redshift,
the majority of SMGs are located in protocluster cores. At all redshifts and flux densities, $\la 20$ percent of SMGs are
located in protocluster outskirts.
\label{fig:smg_frac_env}}
\end{figure*}

We have shown that a random galaxy in a protocluster core is more likely to be submm-bright than in protocluster outskirts or
the field, but it is also worth considering the probability that a random (observationally-identified) SMG is located in a protocluster core.
Field galaxies represent 89 percent of the mock sample, therefore, although protocluster core members are more likely
to be SMGs, it does not necessarily follow that most SMGs at a given redshift are in located in protocluster cores.
To address this question, we quantify the fraction of SMGs located in different environments as a function of $S_{870}$ flux density and redshift in Figure \ref{fig:smg_frac_env}. At all redshifts considered, at low flux densities, most SMGs are located in the
field, whereas above a transition flux density, most SMGs are located in protocluster cores. This transition flux density decreases
with redshift, ranging from $S_{870} \sim 3$ mJy at $z \sim 2$ to $S_{870} \sim 1.5$ mJy at $z \sim 4$.
Protocluster outskirts (as defined in this work, i.e.\, galaxies that will reside in clusters at $z = 0$ but are not in the most massive progenitor halo at the epoch of observation) never dominate the population: at all redshifts and flux densities considered, the fraction of
SMGs in protocluster outskirts is $\la 20$ percent.

We caution that the above details (e.g.\,the transition flux) may be affected by the underprediction of the overall submm number counts; see Section \ref{dis:underprediction}. In follow-up work, we will attempt to recalibrate the model to better reproduce the submm counts and will then revisit these results. In the interim, we can make the crude assumption (which is certainly incorrect in detail but perhaps still reasonable) that such a recalibration would boost the submm flux densities by a constant factor. As seen in Fig. \ref{fig:submm_ncounts}, multiplying the dust masses by a factor of 2.5 (which boosts the flux density by 81 percent) brings the predicted and observed counts into agreement. For convenience, we have plotted these rescaled fluxes on the top $x$-axes in Fig. \ref{fig:smg_frac_env}.

\subsection{Why is there an excess of submm-bright galaxies in protocluster cores?} \label{sec:main-props}

\begin{figure*}[ht!]
\includegraphics[width=\textwidth]{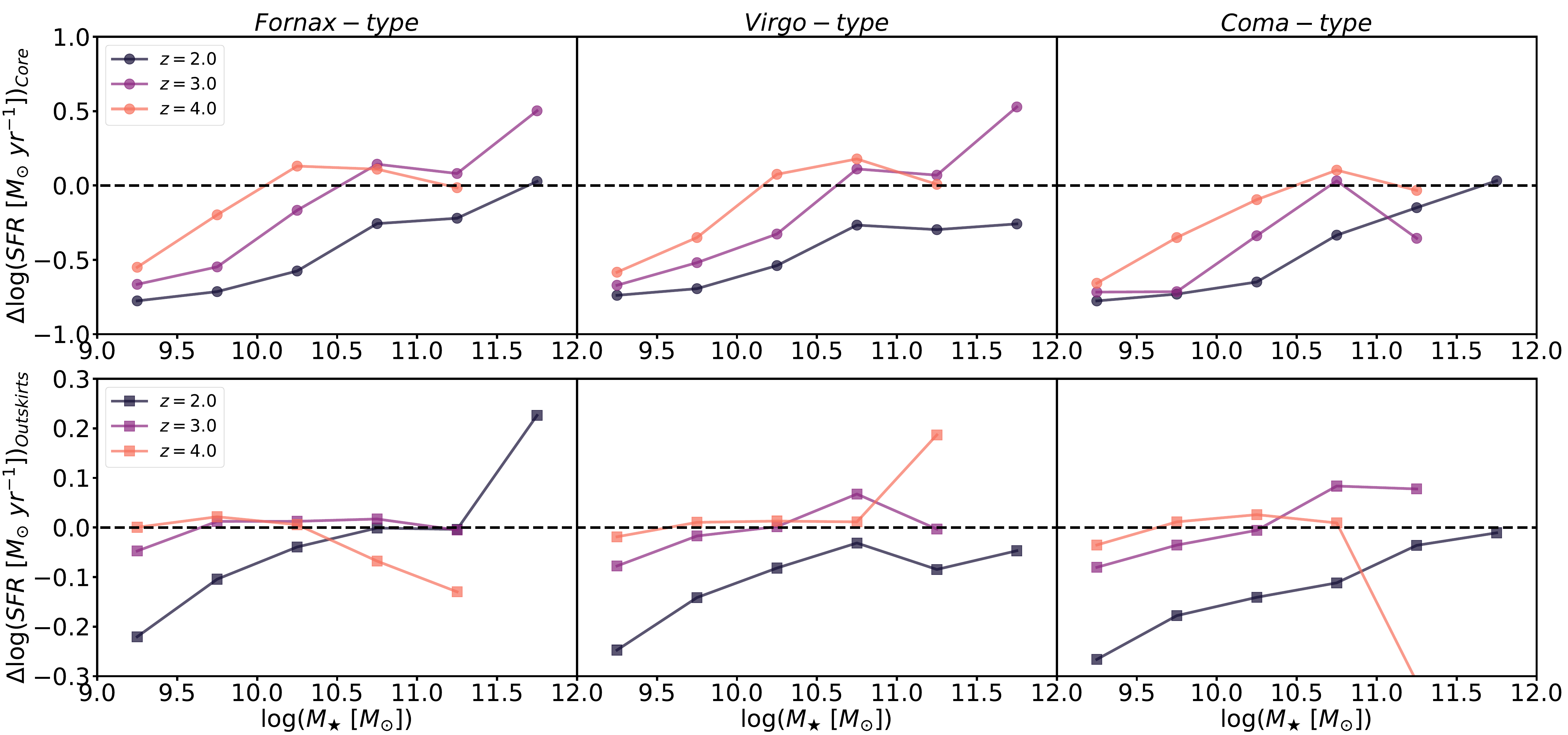}
\caption{The median SFR as a function of stellar mass for galaxies in protocluster cores (\emph{top row}) and in protocluster outskirts (\emph{bottom row}) minus the median SFR of field galaxies (i.e.~the `excess SFR' relative to the field, denoted $\Delta \log({\rm SFR})$) at $z\sim2$, 3, and 4 (\emph{dark blue, purple,} and \emph{orange lines}, respectively) for Fornax, Virgo, and Coma-type protoclusters (\emph{first, second and third columns}, respectively). The \emph{black dashed line} indicates $\Delta \log({\rm SFR}) =$ 0 (median SFR equal to the field value).
Overall, galaxies in protocluster cores and outskirts do not exhibit enhanced SFRs relative to the field, i.e.~`starbursts' are
uncommon. At lower stellar masses, galaxies in protocluster cores have \emph{lower} SFRs than those in the field (by as much as
an order of magnitude); this trend is stronger at lower redshift.
Galaxies in protocluster outskirts have SFRs similar to those of field galaxies except for at $z \sim 2$, where lower-mass galaxies
have SFRs slightly suppressed relative to the field.
\label{fig:delta_sfr}}
\end{figure*}

In this section, we seek to understand the physical reason(s) behind protocluster cores exhibiting a higher submm-bright fraction compared to both protocluster outskirts and the field. As noted above, one potential explanation for the excess of submm-bright sources in protocluster cores is that such environments may have an excess of starbursts due to more frequent interactions and mergers (see
\citealt{casey16} and references therein). 
It is also possible that this excess
is simply due to a relative overabundance of massive galaxies and thus `oversampling' of the high-mass end of the star formation main sequence (SFMS). A third possibility is that galaxies in protocluster cores tend to be
more dust-rich than those in other environments.

To investigate the first possibility, in Figure \ref{fig:delta_sfr}, we quantify the difference between the median SFR of galaxies in the field and those of galaxies in protocluster cores (top) and outskirts (bottom) as a function of stellar mass at $z=2$, 3, and 4. As in previous plots, the different columns correspond to the three protocluster types considered here. For more detail, the full SFR-$M_{\star}$ distributions for the same redshifts and protocluster types are shown in the Appendix (Figure \ref{fig:sfr_smass}).  

From the top row of Figure \ref{fig:delta_sfr}, we see that galaxies in protocluster cores above a threshold
stellar mass tend to lie on or slightly above the SFMS defined by field galaxies. 
Below this threshold stellar mass, core members have SFRs that tend to be lower
than those of field galaxies by as much as $\sim 1$ dex.
The stellar mass threshold depends on redshift and protocluster type. At $z \sim 4$, the stellar mass threshold is
$\log(M_{\star}/\msun) \sim 10.2$. At $z \sim 3$, this value is shifted towards higher stellar masses, $\log(M_{\star}/\msun) \sim 10.7$, and at $z \sim 2$, it
its $\sim 11.7$. At all redshifts considered,
the difference between the median SFRs of field galaxies and those in protocluster cores is greatest at the lowest masses considered.
These results clearly show that the first potential explanation for the excess of SMGs in protocluster cores, that
dense environments feature a greater starburst fraction, does not hold for our model. This is further reinforced by examination
of Figure \ref{fig:sfr_smass} in the Appendix.

For galaxies in protocluster outskirts (bottom row of Figure \ref{fig:delta_sfr}), at $z \sim 3$ and 4, outskirts members typically
lie near the field SFMS. 
At $z \sim 2$, low mass ($\log(M_{\star}/\msun) \lesssim 10.5$) galaxies have SFRs that are suppressed by at most a factor of 2
relative to the field, whereas higher-mass member galaxies have SFRs similar to those of field galaxies. 

\begin{figure*}[ht!]
\includegraphics[width=\textwidth]{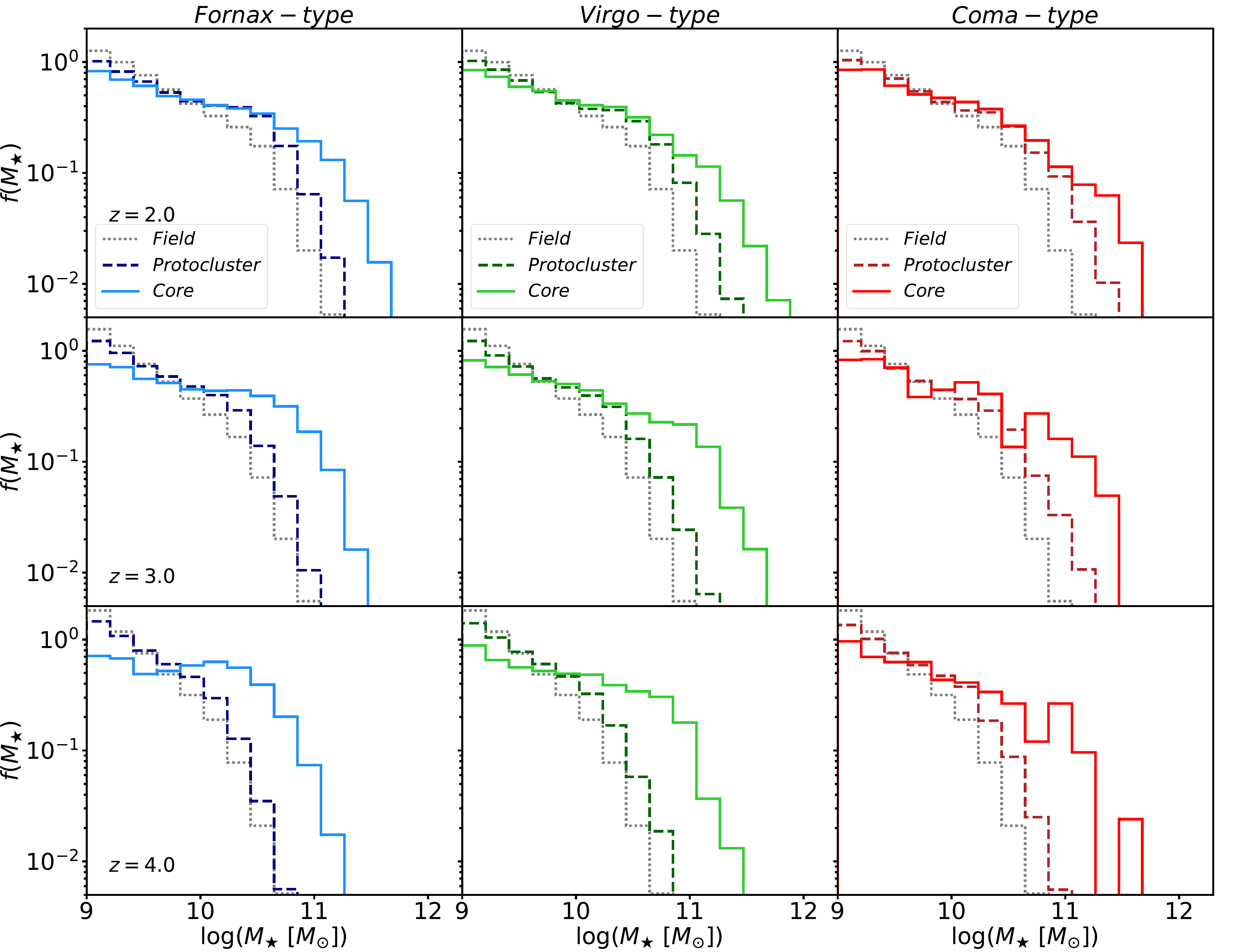}
\caption{Normalised SMFs of galaxies in protocluster cores (\emph{solid lines}), protocluster outskirts (\emph{dashed}), and the field (\emph{grey dotted}) at $z\sim2$, 3, and 4 (\emph{first, second and third rows}, respectively) for Fornax-, Virgo, and Coma-type protoclusters (\emph{first, second, and third columns}, respectively).
In all panels, protocluster cores exhibit a relative excess of high-mass galaxies compared to protocluster outskirts and the field.
The SMFs of protocluster outskirts and the field are similar.
\label{fig:smass_dist}}
\end{figure*}

We now turn to the second possible explanation, that protocluster cores might exhibit an excess of high-mass galaxies relative to
protocluster outskirts and the field.
To examine this hypothesis, we present the stellar mass probability density functions (PDFs) of core, outskirts, and field galaxies at different redshifts ($z\sim2$, 3, and 4; $\Delta z = 0.25$) for the three protocluster types (Fornax-, Virgo-, and Coma-type progenitors) in Figure \ref{fig:smass_dist}.
The most notable trend is that protocluster cores exhibit an excess of high-mass galaxies ($M_{\star} \ga 10^{10.5} \msun$),
i.e. a flatter stellar mass function (SMF), relative to both protocluster outskirts and the field at all redshifts and for all protocluster types.
The difference is more pronounced at higher redshift.
Note that the stellar mass distribution of galaxies in Coma-type protoclusters at $z\sim4$ exhibits a bump at $\log(M_{\star} / \msun) \sim 10.7$. This is likely due to statistical fluctuations owing to the small sample of galaxies with $M_{\star} \ga 10^{10.5} \msun$ in Coma-type protoclusters
at $z \sim 4$. 

The SMFs for the field and protocluster outskirts are similar, though protocluster outskirts exhibit a slightly flatter
SMF than the field. This difference is most pronounced for the outskirts of Coma-type protoclusters
(i.e.\,the highest overdensities on average) and for Virgo-type protoclusters at $z \sim 2$.
Thus, both distributions become different as the redshift decreases.

\begin{figure*}[ht!]
\includegraphics[width=\textwidth]{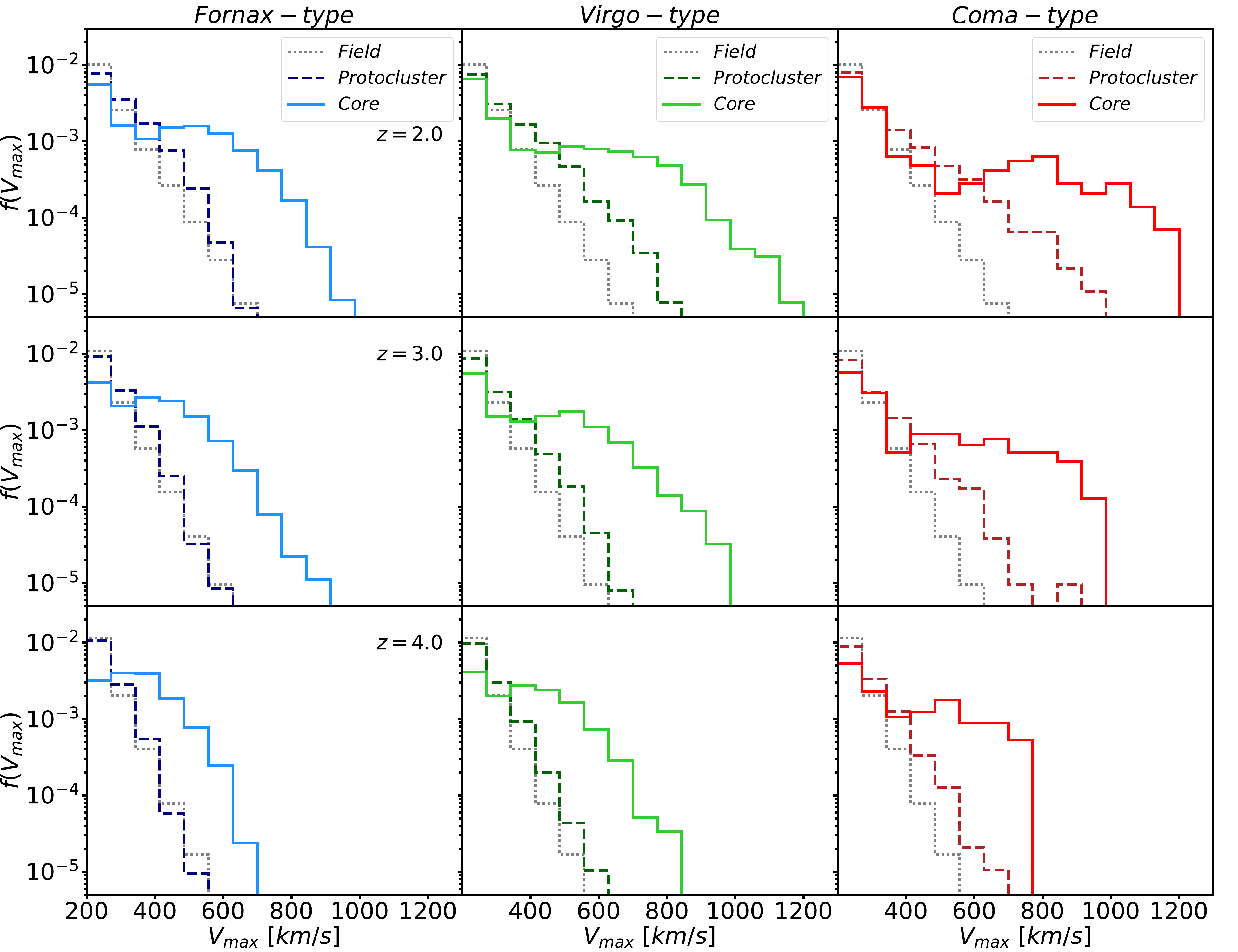}
\caption{Similar to Figure \ref{fig:smass_dist}, but for $v_{\rm max}$ (as a proxy for dark matter halo mass).
Protocluster cores exhibit flatter $v_{\rm max}$ distributions
than both protocluster outskirts and the field. The difference
is greatest for Coma-type protoclusters and increases with decreasing
redshift.
\label{fig:vmax_dist}}
\end{figure*}

What is the origin of this excess of high-mass galaxies in protocluster cores? It may be due to baryonic processes that operate
preferentially in protocluster cores, but it could also be a result of hierarchical growth of dark matter halos.
To examine the latter possibility, in Figure \ref{fig:vmax_dist}, we plot the PDFs of $v_{\rm max}$ values (as a proxy for
halo mass) for galaxies in the three different environments.
We see the same qualitative trends as for the SMF: protocluster cores exhibit flatter $v_{\rm max}$
distributions than both protocluster outskirts and the field, and the difference is more pronounced for Coma-type
protoclusters than for Fornax-type ones (i.e.\,for higher overdensities). The difference increases with decreasing
redshift. For Coma-type protoclusters and Virgo-type ones at $z \sim 2$, the outskirts exhibit slightly flatter $v_{\rm max}$
distributions than the field but are more similar to the field than to protocluster cores. The difference
again increases with decreasing redshift. The results suggest that an excess of high-mass dark matter halos in protocluster
cores results in an excess in high-mass galaxies and thus a higher SMG fraction in cores (as long as the quenched
fraction is low).\footnote{We have confirmed that the stellar-to-halo mass relation at each of the redshifts considered is insensitive to environment.}

\begin{figure*}[ht!]
\includegraphics[width=\textwidth]{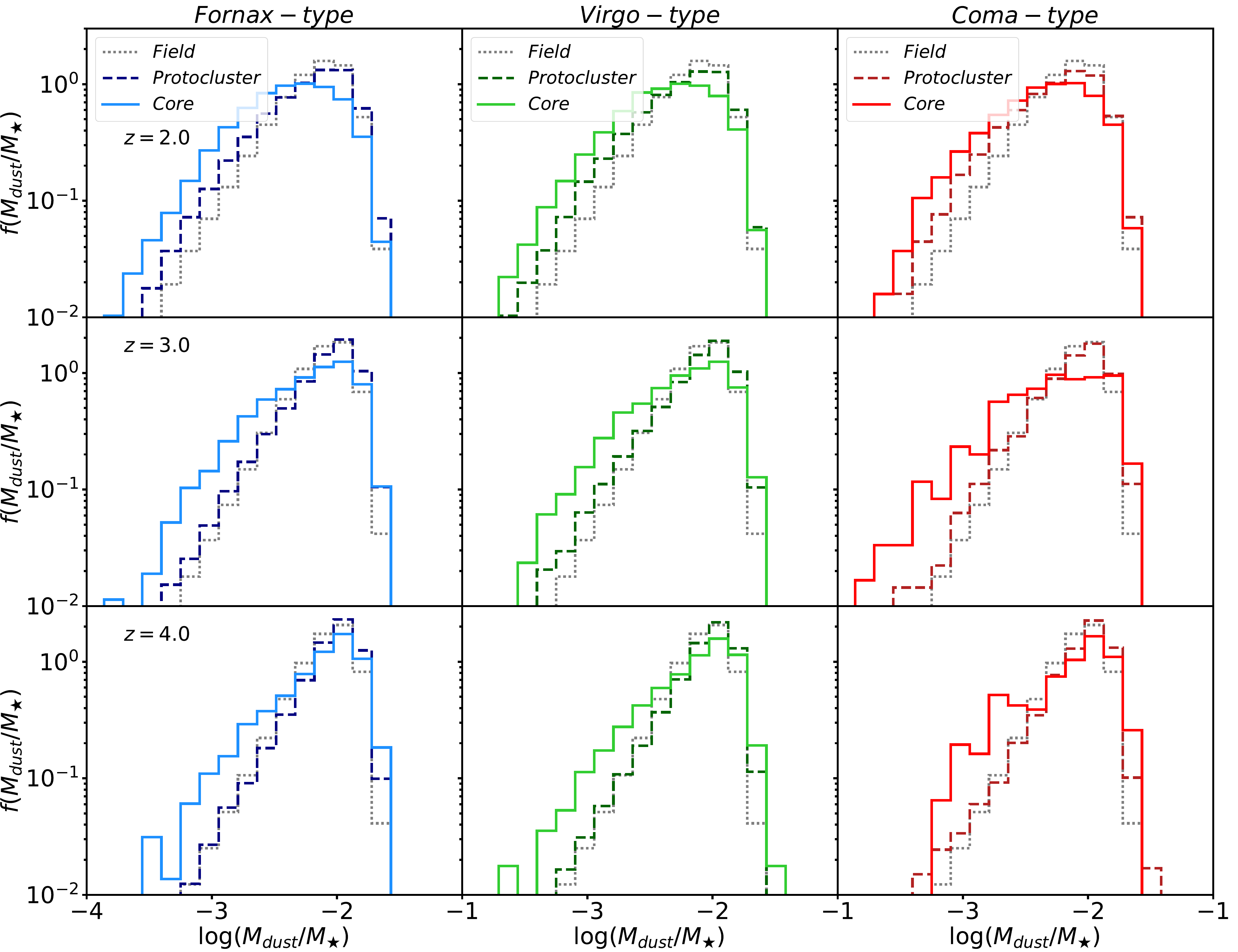}
\caption{Similar to Figure \ref{fig:smass_dist}, but for dust-to-stellar
mass ratio. For all redshifts and protocluster types, the distributions
are similar for the three different environments. They peak at the
expected value of $\sim 0.01$. There is a tail to low dust-to-stellar
mass ratios that is slightly more enhanced in protocluster cores,
but these excess dust-poor sources constitute a small fraction of
the population.
\label{fig:mdust_dist}}
\end{figure*}

Finally, we explore the third possibility, that galaxies in
protocluster cores are more dust-rich than galaxies in protocluster
outskirts and the field (which, for fixed SFR, stellar mass, and
redshift, would yield a colder effective dust temperature and thus
higher submm flux density). In Figure \ref{fig:mdust_dist}, we
plot the PDFs of the dust-to-stellar-mass
ratio for galaxies in protocluster cores, protocluster outskirts,
and the field for the three different protocluster types at
$z \sim 2$, 3, and 4. We see that the distributions are similar,
with a peak at the expected value of $\sim 0.01$. Protocluster
cores exhibit a slightly enhanced tail to \emph{lower} dust-to-stellar
mass ratios, but this tail constitutes a small fraction of the
overall population. We conclude that the dust content is almost
independent of environment, and thus the third potential explanation
does not explain the excess of SMGs in protocluster
cores in our model.

\subsection{Optical counterparts of SMGs} \label{sec:optical-counterpart}

\begin{figure*}[ht!]
\includegraphics[width=\textwidth]{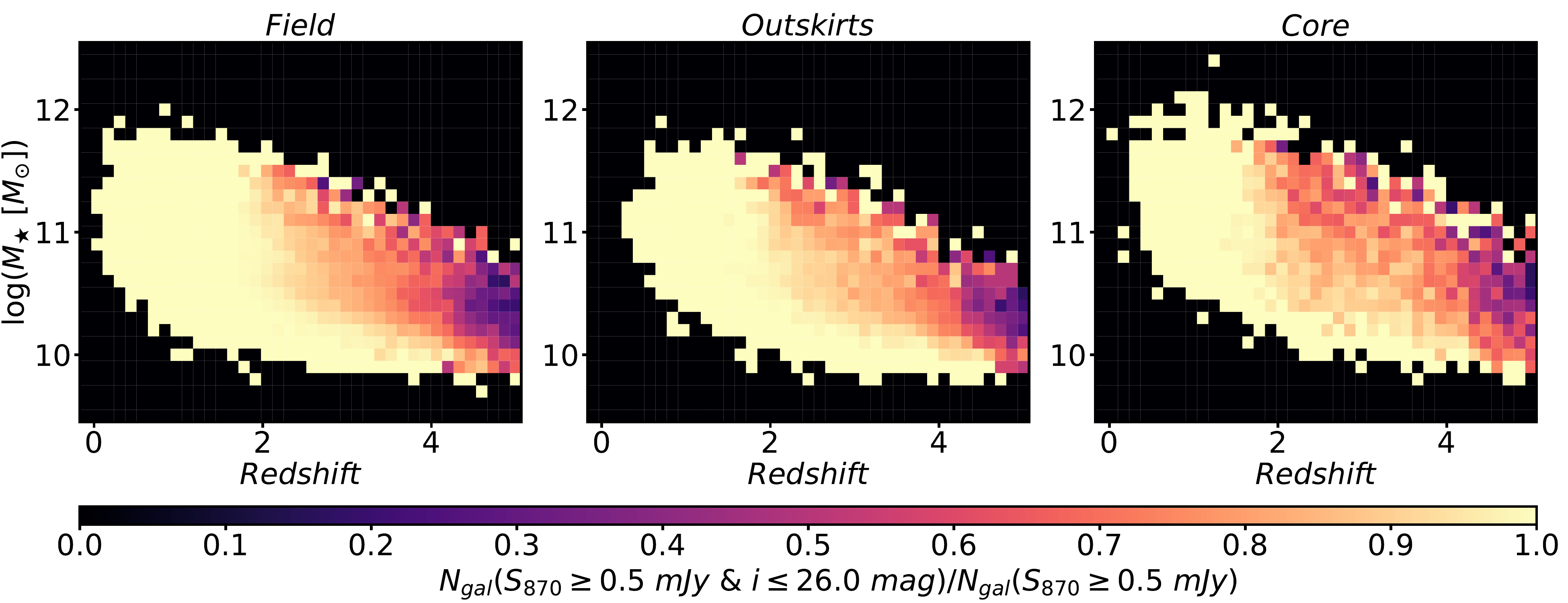}
\caption{The fraction of submm sources ($S_{870} \geq 0.5$ mJy) with optical counterparts ($i \leq$ 26 mag) in 2D bins of redshift and stellar mass for the field (\emph{left}), protocluster outskirts (\emph{middle}), and 
protocluster cores (\emph{right}). At fixed redshift and stellar mass, the fraction of SMGs that have optical counterparts does not exhibit a strong dependence on the environment. However, massive galaxies are generally more difficult to detect in both wavelengths simultaneously,
i.e.\,more-massive SMGs are less likely to have optical counterparts
than lower-mass SMGs.
\label{fig:frac_submm_opt}}
\end{figure*}

We now consider the relationship between submm- and optically selected populations in different environments in our model.
As shown by \citet{rotermund21}, just 4 out of 14 submm sources in the SPT2349-56 protocluster were observed in optical bands. This highlights the difficulties of identifying the most highly dust-attenuated sources at short wavelengths (see \citealt{cochrane21,cochrane24} and recent observational studies concerning optically faint SMGs, e.g.~\citealt{smail21,smail23}). In this section, we discuss the augmentation of our mock catalog with predictions for optical photometry and present the fraction of submm-bright sources in our catalog that would have  counterparts in deep optical imaging.

Typically, for observational data, photometric catalogs are constructed by making aperture measurements at the same positions in multiple bands; these positions are defined by a reference image, which is commonly a $\chi^2$ detection image constructed using several bands (\citealt{szalay99}; see \citealt{mendes19, almeida22, laigle16, weaver22} for examples). On the other hand, the Hyper Suprime Cam Subaru Strategic Program (HSC-SSP; \citealt{aihara18, huang18}) constructs the reference image for source detection using the $i$-band.

A common practice when using optical data is implementing a magnitude limit to avoid contamination of low signal-to-noise objects.
For instance, \citet{toshikawa18} and \citet{rotermund21} selected samples of Lyman Break Galaxies using a $5\sigma$ magnitude limit in the $i$-band (also requiring a $3 \sigma$ detection in the $r$-band), which corresponds to 25.9 mag and 26.2 mag, respectively, for those studies. Here, we select optical sources from our mock catalogs using an $i$-band magnitude limit of $26$ mag, similar to the limit applied in these works.

We analyze the fraction of SMGs ($S_{870} \geq$ 0.5 mJy) that would also be detected in the optical ($i \leq$ 26 mag). In Figure \ref{fig:frac_submm_opt}, we explore how this fraction depends on redshift and stellar mass for galaxies in the field, protocluster outskirts, and protocluster cores. Here, we do not show the distribution separated by protocluster type because we did not find any trend with this property. As expected, the probability of finding an optical counterpart for an SMG decreases with increasing redshift, since distant sources appear fainter. 
At high redshift ($z \gtrsim 3.0$), within a fixed redshift bin, there is an anti-correlation between the fraction of SMGs with optical counterparts and stellar mass, i.e.\,more-massive
SMGs are less likely to have optical counterparts compared to lower-mass SMGs due to dust obscuration increasing with stellar mass (see the next subsection). This trend appears to be independent of the environment. We note that for a given redshift bin
at high redshift ($z \gtrsim 3.0$), the overall optical counterpart probability is slightly lower for protocluster core galaxies compared to those in protocluster outskirts, and the fraction for protocluster outskirts is slightly lower than that for field galaxies. 

\subsection{Fraction of dust-obscured sources} \label{sec:dust-frac}

\begin{figure}[ht!]
\includegraphics[width=\columnwidth]{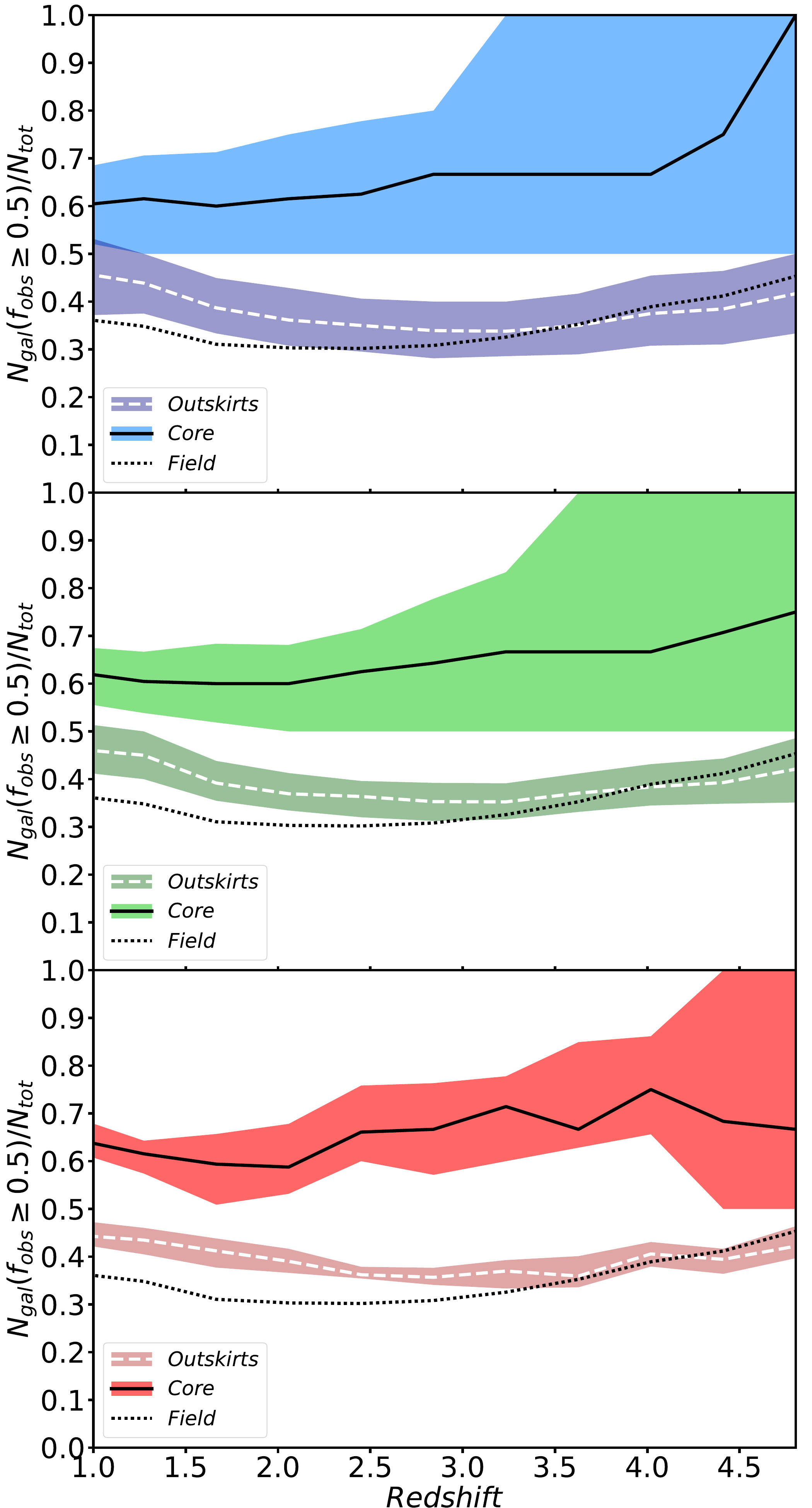}
\caption{The redshift evolution of the fraction of dust-obscured galaxies ($f_{\rm obs} \equiv {\rm SFR_{IR}/SFR_{inst.}} \geq 0.5$) in the three different environments.
Colored regions represent the 25th-75th percentile ranges for protocluster outskirts and cores, while the lines show the median $N_{\rm gal}(f_{\rm obs} \geq 0.5)/N_{\rm tot}$ value in a given redshift bin. We do not find any significant redshift evolution of the fraction of dust-obscured sources in the three environments. Across protocluster types, $\sim 65$ percent of protocluster core galaxies are dust-obscured, while this percentage is $\sim 40$ percent for galaxies in protocluster outskirts and the field.  
\label{fig:frac_fobs}}
\end{figure}
\begin{figure}[ht!]
\includegraphics[width=\columnwidth]{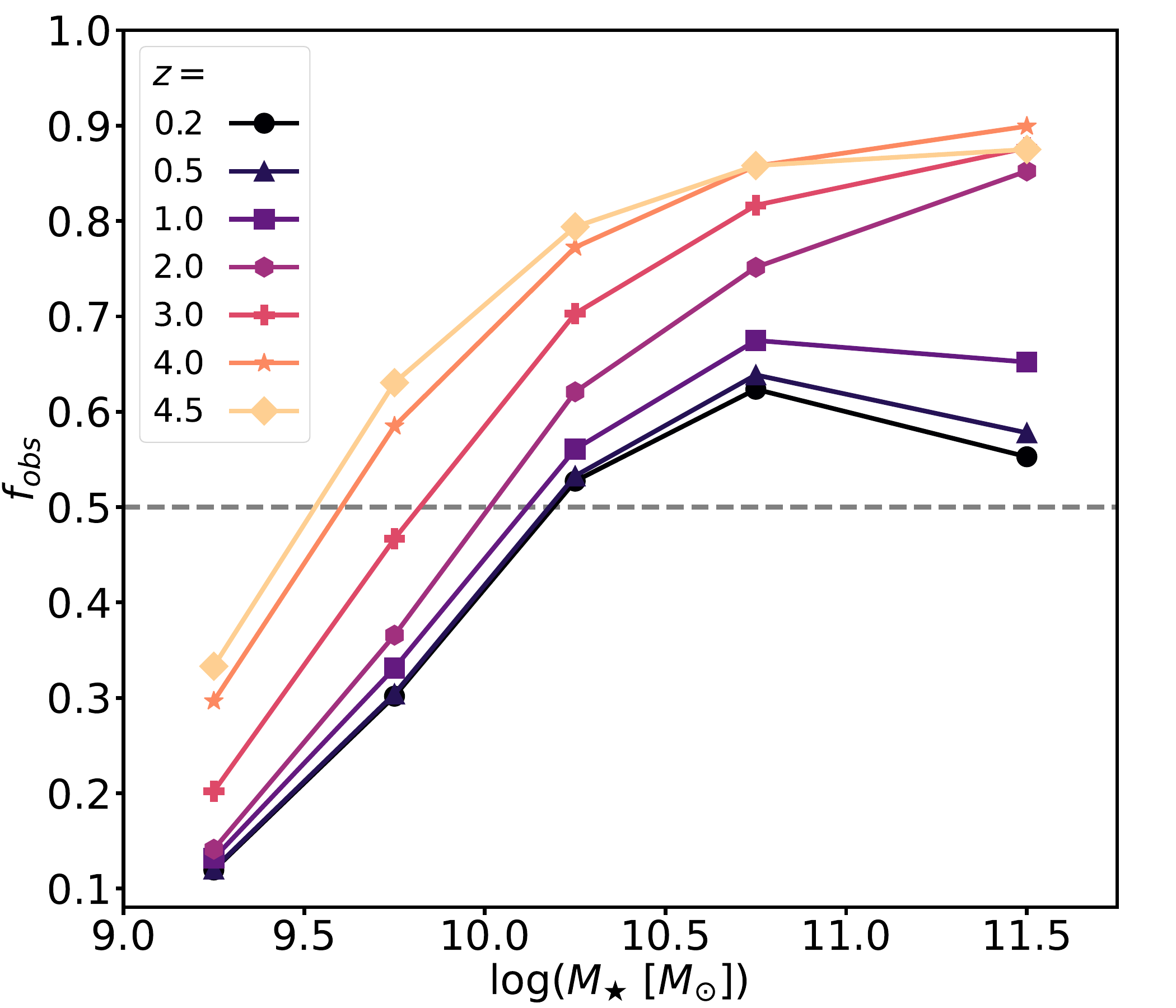}
\caption{The median obscured SFR fraction ($f_{\rm obs} \equiv {\rm SFR_{IR}/SFR} \geq 0.5$) versus stellar mass. The different curves show the relation for seven redshift bins from $z\sim 0.2$ to $z\sim 4.5$. The grey dashed line marks our definition of dust-obscured galaxies (an obscured SFR fraction of greater than 50 percent).
On average, the obscured SFR fraction increases with stellar mass. At fixed stellar mass, the obscured SFR fraction is higher at higher redshifts, but the variation with redshift is subdominant to that with stellar mass.
\label{fig:fobs_smass_z}}
\end{figure}

To understand the above results, it is worthwhile to investigate the fraction of SFR that is obscured.
Following the definition of \citet{zavala21}, we consider dust-obscured sources as all mock galaxies with obscured fraction $f_{\rm obs} \geq 0.5$, where $f_{\rm obs} =$ SFR$_{\rm IR}/$SFR. SFR$_{\rm IR}$ and SFR denote the star formation rate derived from the total galaxy IR luminosity (light absorbed by dust; see Section \ref{sec:templates_app}) and the total star formation
rate, respectively.
To obtain SFR$_{\rm IR}$, we use the conversion of \citet{kennicutt98}: 
\begin{equation}
 {\rm SFR}_{\rm IR} (\msun \ {\rm yr}^{-1}) = 3.0\times 10^{-44}\, L_{\rm IR} ({\rm erg \ s}^{-1}).
\end{equation}
The conversion factor is obtained assuming a \citet{kroupa01} IMF. However, there is no significant difference compared to a \citet{chabrier03} IMF \citep{madau14}.

In Figure \ref{fig:frac_fobs}, we show the fraction of dust-obscured galaxies as a function of redshift for galaxies in protocluster cores, protocluster outskirts, and the field.
The different rows correspond to different protocluster types. In each panel, for protocluster cores and outskirts, lines indicate the medians taken over the population
of protoclusters in the mock, and the shaded regions denote the 25th-75th percentile range. The line for the field indicates the value for the entire field sample at a given
redshift.

Unlike the redshift evolution of the fraction of submm sources (Figure \ref{fig:frac_submm}), the fraction of dust-obscured galaxies does not statistically depend on redshift or protocluster type. However, it is significantly higher in protocluster cores ($N_{\rm gal}(f_{\rm obs} \geq 0.5)/N_{\rm tot} \sim 0.65$) compared to protocluster outskirts and the field. In the latter two environments, the fractions of dust-obscured objects are similar ($N_{\rm gal}(f_{\rm obs} \geq 0.5)/N_{\rm tot} \sim 0.40$).
The reason that the fraction of SMGs depends on redshift but the obscured fraction does not is that the former depends on the absolute values of the SFR and dust mass, whereas the
latter does not. At lower redshifts, there are fewer galaxies with high SFRs ($\ga 100 \msun$ yr$^{-1}$) because at fixed mass, the SFR decreases with decreasing redshift due
to the redshift evolution of the normalization of the SFMS and quenching of high-mass galaxies (`downsizing').

In order to understand the obtained trend in Figure \ref{fig:frac_fobs}, in Figure \ref{fig:fobs_smass_z}, we present the median value of the obscured fraction ($f_{\rm obs}$) versus stellar mass in seven redshift bins from $z = 0.2$ to 4.5. At all redshifts, the obscured fraction increases with stellar mass, as suggested by some observational studies \citep[e.g.][]{garn10,whitaker17}. This trend is also seen in versions of \texttt{L-GALAXIES} with explicit dust modelling, which show a strong correlation between dust-to-gas ratio and stellar mass at all redshifts \citep{Vijayan+19,Yates+24}. Moreover, at fixed mass, galaxies at higher redshifts exhibit higher $f_{\rm obs}$. For instance, the stellar mass threshold above which 50 percent of galaxies are classified as dust-obscured ($f_{\rm obs} \geq 0.5$) is $\log(M_{\star}/\msun) = 10.15$, 10.04, 9.88, 9.67 at $z \sim 1$, 2, 3, and 4, respectively. 

\section{Discussion} \label{sec:discussion}

\subsection{Why do protocluster cores differ from less-overdense regions?} \label{dis:downsizing}

We have shown that in the \texttt{L-GALAXIES} model, galaxies in protocluster cores are more likely to be submm-bright and dust-obscured than galaxies in
protocluster outskirts or the field. Interaction-driven starbursts are not responsible for the
excess of SMGs in protocluster cores. Moreover, there is no clear correlation between environment and inferred dust-to-stellar mass ratios in the model. Instead, the reason that protocluster cores have a higher SMG fraction is simply that they exhibit an excess of high-mass galaxies relative to less-overdense
regions. The submm-bright fraction decreases with decreasing redshift as galaxies at the high-mass end of the SFMS are quenched and thus are
no longer submm-bright.
The relative excess of high-mass galaxies in protocluster cores also results in a higher fraction of dust-obscured galaxies because the fraction of star formation that is obscured increases
with stellar mass. We furthermore showed that protocluster cores have a `top-heavy' dark matter halo mass function relative to less-overdense environments.

Together, these results imply that in our model, the excess of SMGs in protocluster cores is driven by hierarchical growth of dark matter substructure, not baryonic processes.
At fixed redshift, many properties of galaxies (e.g.\,gas, dust, and black hole masses; SFRs; metallicities) scale with stellar mass, which, in turn,
scales with halo mass. Thus, if the halo mass function depends on environment, so too will distributions of the aforementioned galaxy properties
\emph{even if all baryonic processes are independent of environment}. Put otherwise, even if a galaxy `knows' only about its own dark matter halo,
the galaxy population can still exhibit significant environmental dependencies purely due to the dependence of the halo mass function on environment.
These environmental dependencies may give a false impression that processes such as merger-induced starbursts, tidal interactions, and ram pressure stripping
are important when in reality the dark matter `backbone' alone causes the differences.

There is some observational support for the above picture. First, we will discuss whether there is observational evidence
that suggests protoclusters have a greater starburst fraction than the field.
\cite{wang2016} find an elevated starburst fraction in a $z = 2.51$ X-ray-detected cluster. \citet{hayashi16} demonstrated
that in the $z \approx 2.5$ protocluster USS 1558-003, massive galaxies
lie along the SFMS, whereas lower-mass galaxies have elevated SFRs.
\citet{shimakawa18a} show an enhancement in the SFR-stellar mass relation of HAEs in the most
overdense regions of a $z = 2.51$ protocluster.
\citet{lemaux22} find that at $2 < z < 5$, the average SFR increases with galaxy overdensity and that this trend
is primarily driven by a relative excess of high-mass in more-overdense environments. \citet{perez-martinez23} show
that the SFRs of members of the $z = 2.16$ Spiderweb protocluster are consistent
with the SFMS. SPT2349-56 \citep{strandet17,miller18,hill22} and the `Distant Red Core' (DRC) \citep{oteo18,ivison20} are considered
two `poster child' high-redshift protocluster cores that are rich in SMGs. In both cases, the member galaxies lie near the SFMS, i.e.\,there is no evidence
that the member SMGs are starbursts \citep{rotermund21,long20}.
Given the above, there is not
yet a clear observational consensus
about whether SMGs in protoclusters
tend to lie about the SFMS.

What do observations say about the SMF in protoclusters?
Studies of the SMF in lower-redshift clusters have yielded conflicting conclusions \citep[e.g.][]{vulcani13,vanderburg2013,vanderburg2018,vanderburg20,annunziatella14,davidzon16,tomczak17}. 
\citet{shimakawa18a,shimakawa18b} analysed HAEs in a sample of $z \sim 2$ protoclusters, finding top-heavy SMFs
in the most overdense subregions of the protoclusters.
\citet{edward24} analysed the SMFs of 14 protoclusters at $2.0 < z < 2.5$ using data from the COSMOS2020
catalog \citep{weaver22}. They found that the SMFs of star-forming galaxies do not differ from the field SMF,
whereas quiescent galaxies exhibit a flatter SMF than in the field.
\citet{forrest24} find an excess of high-mass galaxies in overdense environments at $z \sim 3.3$.
JWST observations should help better characterise the SMF in (proto)clusters and resolve this debate. For example,
based on JWST data, \citet{Sun2024} show that the $z = 2.51$ cluster CLJ1001 exhibits an excess of massive galaxies in the
cluster core.

What about simulations?
\citet{Tonnesen2015} investigated the environmental dependence of the stellar mass-halo mass relation in different environments
using cosmological hydrodynamical simulations. They found that the stellar-to-halo mass ratios of central galaxies in
overdense environments were higher than those in less-dense environments. They suggested that the difference is due to
earlier formation times, more frequent interactions at early times, and a greater number of filaments fueling centrals
in overdense environments. \citet{ahad2024} noted that although $z \sim 1.5$ clusters have higher fractions of quenched
galaxies than the field,
massive quenched galaxies around this redshift have ages that appear to be independent of environment. To understand
this puzzle, they investigated the halo properties of massive quenched galaxies in two cosmological hydrodynamical simulations.
They found that the distribution of $v_{\rm max}$ for cluster members is skewed to higher values relative to that for field galaxies,
similar to the results presented in this work. They conclude that secular processes drive the environmental excess of massive quenched
galaxies in high-redshift (proto-)clusters, analogous to what we have argued for the excess of submm-bright galaxies in protocluster
cores.

Further comparisons of the SFR-stellar mass relations, SMFs, and dust-to-stellar mass ratios
between observed protoclusters and the field will help test the physical picture we have proposed and deepen our understanding
of the role of environment in protoclusters.

\subsection{Implications for using SMGs as tracers of protoclusters} \label{dis:tracers}

SMGs are commonly used as beacons of protoclusters \citep[e.g.][]{chapman01,chapman09,daddi09,capak11,dannerbauer14,casey15,casey16,wang2016,wang21,miller18,oteo18,harikane19,gao22,calvi23,zhou24}.
They are intrinsically very luminous, and due to the negative \emph{k}-correction
in the submm, a given SMG is as easily detected in the epoch of reionisation as it is at $z = 1$. These aspects make them useful tracers.
Moreover, their high masses imply that SMGs are preferentially associated with overdensities. Indeed, we see that this is the case in our model:~Figure \ref{fig:smg_frac_env}
shows that the majority of bright SMGs reside in protocluster cores.
This plot implies that if one finds even a single bright SMG, it is likely that one has
identified a protocluster in the sense that the $z = 0$ descendant of the SMG will reside in a cluster (see also \citealt{miller20}).\footnote{We caution that
the exact threshold above which SMGs
are preferentially located in protoclusters may be affected by our model's underprediction of the submm number counts; see Section \ref{dis:underprediction}
for details.}

SPT0311-58 \citep{strandet17,marrone2018,spilker22} is an example of this phenomenon:~initial ALMA observations resolved this
$z = 6.9$ source to be a merger of two galaxies, one of which is a bright SMG ($S_{870} \approx 15.9$ mJy; the companion has 
$S_{870} \approx 2.9$ mJy). The inferred dark matter halo mass suggested that this is one of the highest-mass
halos expected under LCDM, thus representing a very rare peak in the matter distribution \citep{marrone2018}.
\citet{wang21} found that the surrounding field hosts an excess of SMGs.
Recent JWST integral field unit observations revealed ten additional galaxies within a (17 kpc)$^2$ field of view
\citep{arribas2023}. Taken together, these data suggest that SPT0311-58 resides in a protocluster.

However, we stress that even if an SMG is in a protocluster according to our theoretical definition,
it does not necessarily follow that the SMG is located in an overdensity \emph{at the epoch of observation}.
Whether one should see an overdensity around bright SMGs will depend on the redshift, the flux of the SMG,
and the tracer employed (and potentially other factors). Understanding the connection between SMGs and overdensities
of various galaxy populations will require careful modeling. We will investigate this topic in detail in future work.

Our model suggests that bright SMGs are an effective means to identify protoclusters in terms of purity. But, what about completeness? I.e.~what fraction of
protoclusters host SMGs? We have not investigated this question quantitatively using the present model because the answer would certainly be affected
by the underprediction of the submm number counts. Nevertheless, the model yields some insight that should be robust. We return to Figure \ref{fig:frac_submm},
which shows the fraction of galaxies that are submm-bright versus redshift in different environments. Recall that the submm-bright fraction is significantly elevated
in protocluster cores at high redshift, but the enhancement decreases with decreasing redshift due to `downsizing', i.e.~the mass above which
galaxies are predominantly no longer star-forming decreases with decreasing redshift. This suggests that SMGs are increasingly incomplete
tracers of protoclusters at lower redshift. This was previously argued by \citet{miller15}, who applied a semi-empirical model in which the galaxy properties are
determined solely by halo mass and redshift to an N-body simulation. We find that this conclusion holds in a SAM that includes
models for environmentally dependent processes (e.g.~mergers, ram pressure stripping).

\subsection{Sensitivity of results to the overall underprediction of the submm number counts} \label{dis:underprediction}

As shown in Figure \ref{fig:submm_ncounts}, our model underpredicts the overall submm number counts by a factor of a few. Historically, theoretical
models that assume a standard IMF have underpredicted submm number counts \citep[e.g.][]{granato00,benson03,lacey16,somerville12,baugh05,hayward13,hayward21,lovell21},
although the magnitude of the discrepancy has decreased considerably as models have been improved.
To the best of our knowledge, no model can simultaneously
match both the observed submm number counts \emph{and} the number density of massive quenched galaxies \citep{hayward21}.
If the submm counts predicted by our model were consistent with observations, some of the results presented here would be affected.
Importantly, for a fixed flux density threshold, the fraction of SMGs would increase.

However, many of our results
are likely insensitive to this underprediction. Given that the differences between protocluster cores and other environments are primarily driven
by differences in the halo mass function rather than baryonic processes, these differences should be robust to any changes to model parameters
that would be necessary to reproduce the total submm counts.
Moreover, if the underprediction is associated with the treatment of merger-induced starbursts, it is likely that the submm-bright fractions for
protocluster cores would be boosted more than those for the field and protocluster outskirts, further enhancing the difference
between cores and other environments.

Still, it would of course be better if our model predictions better agreed with the observed submm number counts. In subsequent work, we will explore whether
it is possible to simultaneously match the submm number counts and number density of massive quenched galaxies with \texttt{L-GALAXIES}
by using parameter values different than those employed here. Should we succeed, we will be able to revisit the analysis presented here
and check whether our primarily results are indeed robust.

\section{Summary} \label{sec:summary}

Using a mock catalog constructed by applying the \citet{henriques15} version of the \texttt{L-GALAXIES} SAM to the \texttt{Millennium} simulation and then employing the \citet{cochrane23} scaling relations to predict submm
flux densities, we have investigated the relationship between SMGs and galaxy protoclusters. Our main results are the following:  
\begin{enumerate}
    \item By exploring the dependencies on redshift, environment, and protocluster descendant mass, we found that protocluster core galaxies experience a submm-bright phase, whose peak and duration correlate with the $z=0$ mass and flux density limit (Figure \ref{fig:frac_submm}). At $z \ga 2$, protocluster cores exhibit a higher submm-bright fraction than do
    galaxies in protocluster outskirts or the field.
    
    \item We show that the primary driver of this excess of SMGs in protocluster cores is differences in the halo mass functions in different environments: protocluster cores exhibit a more `top-heavy'
     halo mass function (Figure \ref{fig:vmax_dist}), which results in an excess of massive star-forming galaxies relative to other environments (Figure \ref{fig:smass_dist}).
     The excess of SMGs in protocluster cores is not due to a higher starburst fraction (Figure \ref{fig:delta_sfr}) or differences in the dust contents of galaxies in different environments
     (Figure \ref{fig:mdust_dist}), as protocluster cores are similar to other environments in terms of these two properties.
    
    \item Most SMGs with $S_{870} \geq 0.5$ mJy are in the field rather than protoclusters, but bright SMGs are preferentially located in protocluster cores (Figure \ref{fig:smg_frac_env}).
    
    \item Independent of environment, the fraction of SMGs that have optical counterparts decreases with increasing redshift and with stellar mass (Figure \ref{fig:frac_submm_opt}).
    
    \item Our model predicts a higher fraction of dust-obscured sources in protocluster cores compared to protocluster outskirts and the field independent of redshift.
    (Figure \ref{fig:frac_fobs}). The obscured fraction increases strongly with stellar mass and more weakly with redshift (Figure \ref{fig:fobs_smass_z}).
    
\end{enumerate}

Our results add to our understanding of the connection between SMGs and protoclusters. Our mock catalog can be used to optimize the identification, characterization, and interpretation of protoclusters in existing and future surveys.

\section*{Acknowledgements}
We thank the anonymous referee and Ian Smail for the valuable comments.
This work was initiated during the CCA Pre-doctoral Program.
PA-A thanks the Coordenaç\~o de Aperfeiçoamento de Pessoal de Nível Superior – Brasil (CAPES), for supporting his PhD scholarship (project 88882.332909/2020-01). RKC was funded by support for program \#02321, provided by NASA through a grant from the Space Telescope Science Institute, which is operated by the Association of Universities for Research in Astronomy, Inc., under NASA contract NAS 5-03127. RKC is grateful for support from the Leverhulme Trust via the Leverhulme Early Career Fellowship. The Flatiron Institute is supported by the Simons Foundation. MCV acknowledges the Fundaç\~ao de Amparo à Pesquisa do Estado de S\~ao Paulo (FAPESP) for supporting his PhD. LSJ acknowledges the support from CNPq (308994/2021-3)  and FAPESP (2011/51680-6). DR is supported by the Simons Foundation.
\\

\vspace{5mm}

\software{astropy \citep{astropy22},  L-Galaxies \citep{henriques15}}

\appendix

\section{SFR -- stellar mass relation} \label{ap:sfr-smass}
\begin{figure*}[ht!]
\includegraphics[width=\textwidth]{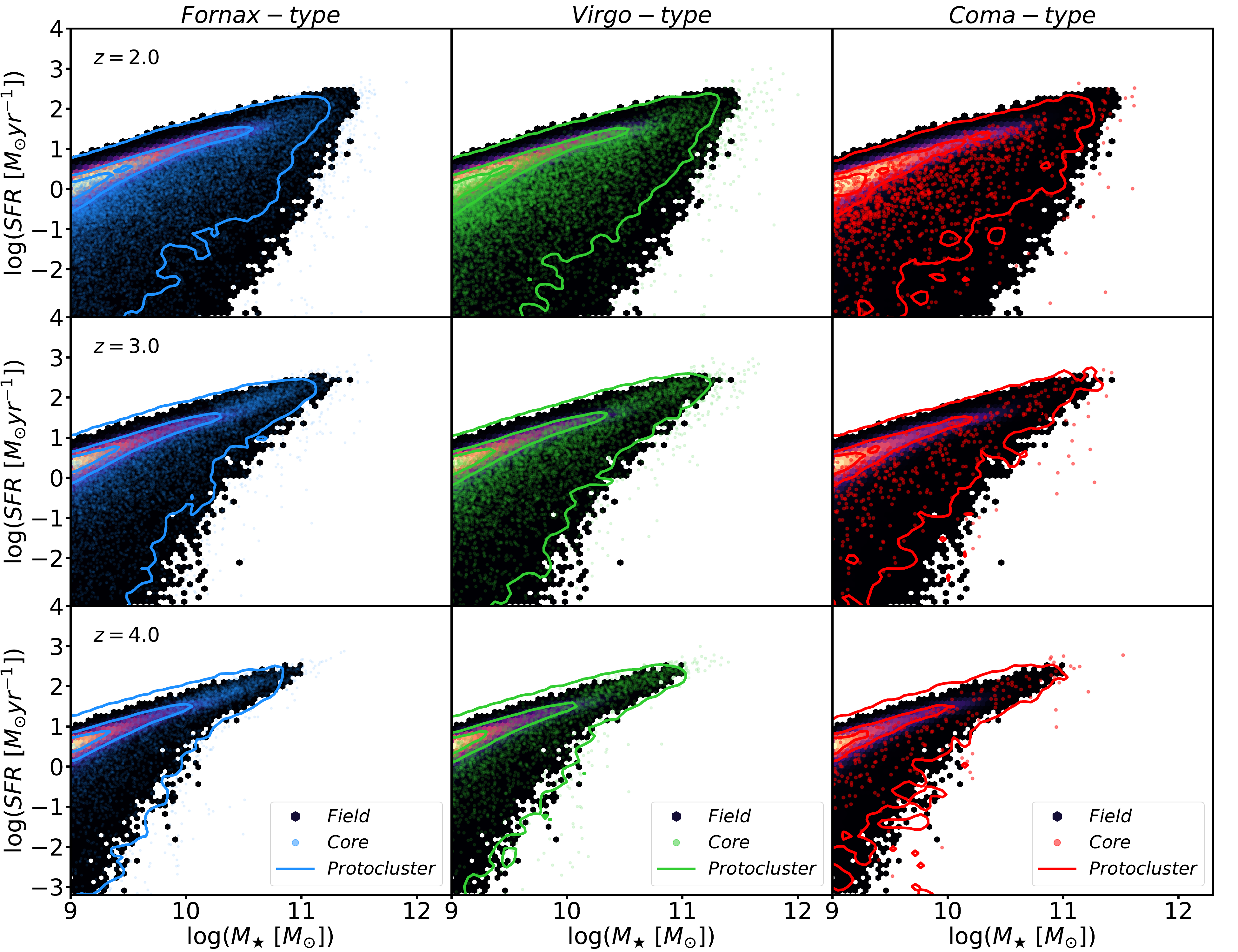}
\caption{SFR versus stellar mass relation for galaxies in protocluster cores (points), protocluster outskirts (contours), and the field (colored hexbin maps)
at $z\sim2.0$, 3.0, and 4.0 (\emph{first, second and third rows}, respectively) for Fornax-, Virgo, and Coma-type protoclusters (\emph{first, second and third columns}, respectively).
For all redshifts and protocluster types, the locus of galaxies in the SFR--stellar mass plane (i.e.~the SFMS) is independent
of environment. In all panels, the colored points at the high-mass tip of the SFMS indicate that protocluster cores host
very high-mass star-forming galaxies that are not represented in the less-dense environments.
\label{fig:sfr_smass}}
\end{figure*}

Figure \ref{fig:sfr_smass} shows the relationship between SFR and stellar mass for galaxies in the three different environments. 
We see that for all redshifts and protocluster types, the distributions of galaxies in this plane are independent of environment
(as already indicated by Figure \ref{fig:delta_sfr}). In particular, galaxies in protocluster outskirts and cores seem to
follow the same (redshift-dependent) SFMS as field galaxies. The most notable difference is that in each panel, there are colored
points at the high-mass tip of the SFMS beyond the region populated by field and protocluster outskirts galaxies. These indicate
that protocluster cores host very high-mass star-forming galaxies that are not found in other environments.

\bibliography{sample631}
\bibliographystyle{aasjournal}

\end{document}